\documentclass{jfm}

\usepackage{graphicx}
\usepackage{newtxtext}
\usepackage{newtxmath}
\usepackage{natbib}
\usepackage{hyperref}
\hypersetup{
    colorlinks = true,
    urlcolor   = blue,
    citecolor  = black,
}

\newcommand{\RomanNumeralCaps}[1]
\linenumbers

\title{Scale-dependent anisotropy, energy transfer and intermittency in bubble-laden turbulent flows}

\author{Tian Ma\aff{1,3}
	\corresp{\email{tian.ma@hzdr.de}},
	Bernhard Ott\aff{2},
	Jochen Fr\"{o}hlich\aff{2}
	\and Andrew D. Bragg\aff{1}
    \corresp{\email{andrew.bragg@duke.edu}}}

\affiliation{\aff{1}Department of Civil and Environmental Engineering, Duke University, Durham, NC 27708, USA
	\aff{2}Technische Universit\"{a}t Dresden, Institute of Fluid Mechanics, 01062 Dresden, Germany
	\aff{3}Helmholtz-Zentrum Dresden -- Rossendorf, Institute of Fluid Dynamics, 01328 Dresden, Germany}

\begin{document}
\maketitle

\begin{abstract}
Data from Direct Numerical Simulations of disperse bubbly flows in a vertical channel are used to study the effect of the bubbles on the carrier-phase turbulence. A new method is developed, based on the barycentric map approach, that allows to quantify the anisotropy and componentiality of the flow at any scale. Using this the bubbles are found to significantly enhance flow anisotropy at all scales compared with the unladen case, and for some bubble cases, very strong anisotropy persists down to the smallest flow scales. The strongest anisotropy observed was for the cases involving small bubbles. Concerning the inter-scale energy transfer, our results indicate that for the bubble-laden cases, the energy transfer is from large to small scales, just as for the unladen case. However, there is evidence of an upscale transfer when considering the transfer of energy associated with particular components of the velocity field. Although the direction of the energy transfer is the same with and without the bubbles, the transfer is much stronger for the bubble-laden cases, suggesting that the bubbles play a strong role in enhancing the activity of the nonlinear term in the flow. The normalized forms of the fourth and sixth-order structure functions are also considered, and reveal that the introduction of bubbles into the flow strongly enhances intermittency in the dissipation range, but suppresses it at larger scales. This strong enhancement of the dissipation scale intermittency has significant implications for understanding how the bubbles might modify the mixing properties of turbulent flows.
\end{abstract}


\section{Introduction}\label{sec: introduction}

Turbulence and multiphase flows are two of the most challenging topics in fluid mechanics and when combined they pose a formidable challenge, even in the dilute dispersed regime \citep{2010_Balachandar}. The focus here is on liquid flows laden with disperse bubbles, which can be particularly challenging since the bubbles can strongly alter the liquid phase turbulence \citep{2005_Mudde,2018_Lohse,2019_Elghobashi}. In particular, the bubbles can modify the turbulence due to production effects arising from the bubble wakes \citep{2010_Riboux,2019_Lai}, enhanced local turbulent kinetic energy dissipation rates in the vicinity of the bubble surfaces \citep{2016_Santarelli_b,2021_Masuk}, and modulation of the liquid mean velocity profile due to interphase momentum transfer, resulting in an alteration of shear-induced turbulence \citep{2013_Lu,2019_Cluzeau,2020_Cifani}. \cite{2020_Mathai} highlighted particular ways in which the classical scenario for single-phase turbulence, based on single-point statistical analysis, is modified due to the bubbles moving relative to the fluid. Turbulence arising from this relative motion is often referred to as bubble-induced turbulence (BIT) and its effects can be captured in the Reynolds-averaged Navier-Stokes (RANS) modelling framework through the inclusion of additional source terms in the relevant transport equations \citep{2014_Fox,2015_Joshi,2020_Ma}.

While significant progress has been made in understanding and characterizing how the bubbles influence the single-point turbulence statistics of the liquid phase, less attention has been paid to the influence of the bubbles on the multiscale/multipoint flow statistics. Those that have considered this aspect have only focused on the kinetic energy spectrum of the liquid velocity fluctuations, with a key observation being that in BIT dominated flows, a power-law behavior for the energy spectrum arises with exponent $-3$ in both the wavenumber or frequency domain \citep{1991_Lance,2011_Roghair,2013_Mendez,2017_Ma}. This behavior was also reported in \cite{2020_Pandey}, who investigated the energy budget equations in wavenumber space and explained the $-3$ slope as arising due to a balance between kinetic energy production due to the bubbles and viscous dissipation. In their analysis they evaluated for the first time (to our knowledge) the nonlinear scale-to-scale energy flux term for bubbly flows, showing a forward (downscale) energy cascade, just as also occurs for single-phase turbulence in three dimensions. An issue with their analysis, however, is that they included the values of the flow at grid points occupied by the bubbles when evaluating the statistics of the carrier phase, thereby contaminating the fluid statistics. 

There have been very few studies exploring the multiscale properties of bubble-laden turbulent flows in physical space, e.g. using structure function analysis. \cite{2005_Rensen} performed hot-film anemometry measurements in the Twente water tunnel and computed the longitudinal second- and fourth-order structure function with the aid of Taylor’s hypothesis. They found an increase of the second-order structure function for the two-phase case compared with the single-phase case under the same bulk Reynolds number, and that this increase was more pronounced at the small scales than the large scales. Their fourth-order structure function results revealed an increase of the intermittency at the small scales of the flow when the flow contained bubbles, even for a relatively low gas void fraction ($0.5\%$). Similar behavior was also observed in \cite{2012_Biferale} when comparing the small-scale properties of boiling and non-boiling convective turbulent flows.

An important aspect yet to be quantified is how the bubbles affect the anisotropy of the flow at different scales. For single-phase turbulence, the energy containing scales in many flows such as those with shear, rotation, and buoyancy are anisotropic \citep{2005_Biferale}. Phenomenological theories of turbulence predict a return-to-isotropy at small enough scales \citep{1941_Kolmogorov_a,1995_Frisch,1997_Sreenivasan}. However, measurements have revealed persistent small-scale anisotropy \citep{1995_Pumir,2000_Shen,2006_Ouellette,2017_Carter}. In contrast to single-phase flow, where energy is often injected into the flow at large scales, bubbles can inject energy into the flow at the scale of their size, which usually corresponds to the small-scales of the turbulence. Since the bubbles have a preferential direction of motion due to buoyancy, this could lead to the injection of strong anisotropy into the flow at the small scales, leading to strong departures from the behavior of the single-phase case. The study of \cite{2020_Pandey} was based on Fourier-space analysis with averaging over spherical shells in wavevector space, and so did not permit them to explore the anisotropy of the flow at different scales.

Another important point is that in \cite{2020_Pandey} the flow had no background turbulence (i.e. all the turbulence was generated by the bubbles), and hence it was not possible to consider how the bubbles modify the turbulence compared with the single-phase case. In order to more fully understand how the bubbles modify the properties of the turbulence, it is desirable to consider a configuration in which the unladen flow is already turbulence, and then one can explore how the bubbles modify the properties of the turbulence when they are introduced.

In the present work we seek to advance the understanding of the properties of bubble-laden turbulent flows across its range of scales. To do this, data from Direct Numerical Simulation (DNS) of finite-size bubbles in a turbulent channel flow with a prescribed bulk Reynolds number is utilized, for different bubble sizes and for mono and bidisperse cases. A new method is developed based on the barycentric map \citep{2007_Banerjee}, and this is used together with the DNS data to quantify the anisotropy of the bubble-laden turbulence flow across the range of its scales. By computing the structure functions of various orders, the direction-dependent liquid velocity fluctuations are also explored at different scales, as is the scale-to-scale energy transfer, and intermittency. These results provide new insights into the properties of bubble-laden turbulent flows, and how they differ from the single-phase counterpart at different scales in the flow.

\section{Direct Numerical Simulations}\label{sec: DNS database}

\subsection{Database}

The DNS data we use is from the studies \citep{2015_Santarelli,2016_Santarelli} that simulated the motion of many thousands of bubbles at low E\"{o}tv\"{o}s number in a vertical turbulent channel flow. The bubbles are handled using an Immersed Boundary Method, and are modeled as rigid spherical objects with a no-slip condition enforced at their surface, representing the behaviour of air bubbles rising in contaminated water. Compared to other simulations of this type \citep[see the related references in][]{2020_Mathai}, these simulations are substantially closer to applications in that they involve a turbulent background flow, contaminated fluid, realistic density ratio, higher bubble Reynolds number, a much larger domain, and a much larger number of bubbles. 

\begin{figure}
	\centering
	\includegraphics[height=5cm]{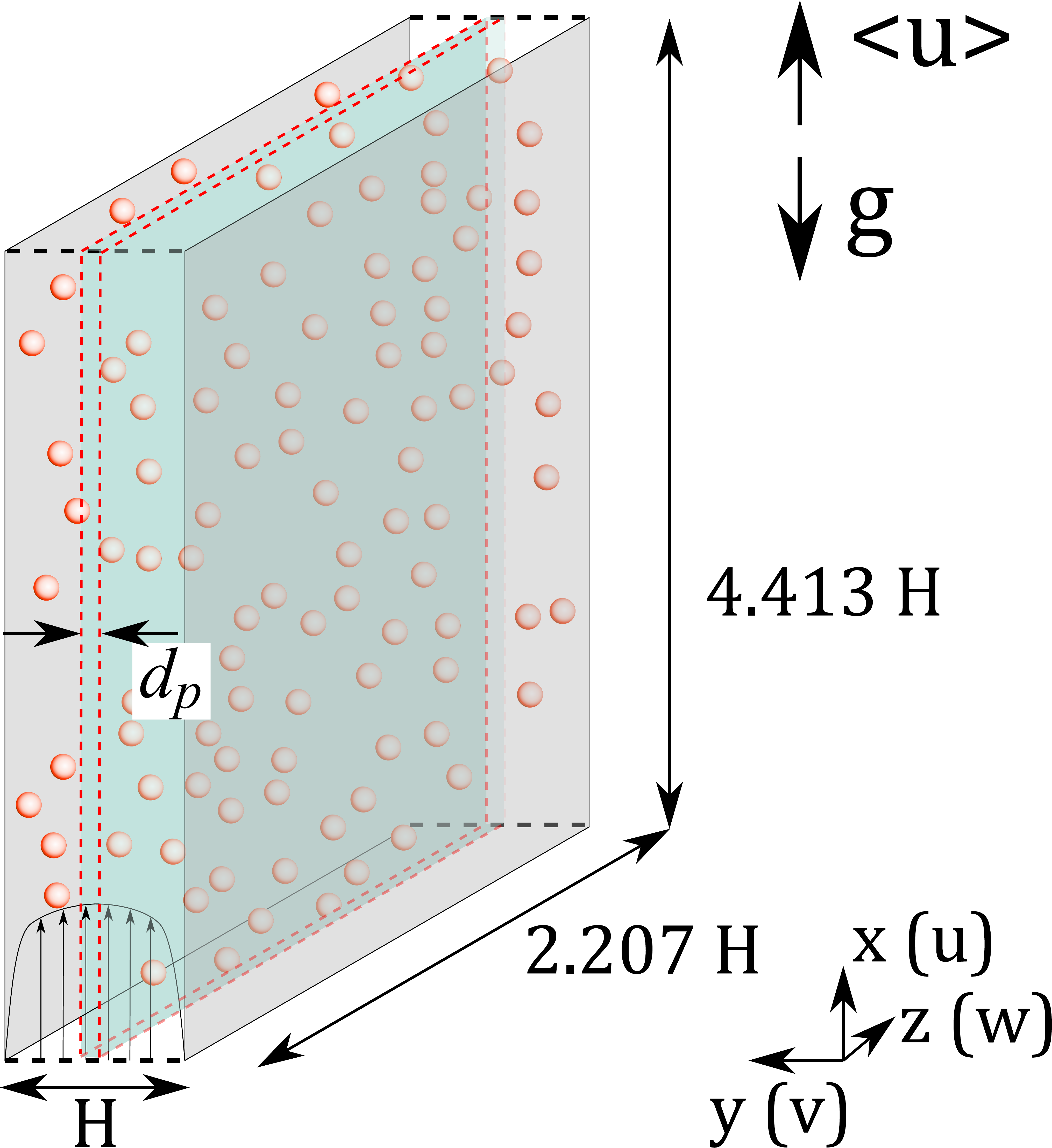}
	\caption{Schematic representation of the DNS configuration (not to scale). The marked region shows the location where the structure functions were calculated.}\label{fig: DNS schema}
\end{figure}

\begin{table}
	\begin{center}
		\def~{\hphantom{0}}
		\begin{tabular}{cccccccc}
			Parameter  &\textsl{Unladen}&\textsl{SmMany}&\textsl{SmFew}&\textsl{LaMany}&\textsl{BiDisp(Sm)}&\textsl{BiDisp(La)}\\
			$N_p$      &--&2880&384&913&1440&546\\
			$\alpha_b$   &--&2.14\%&0.29\%&2.14\%&1.07\%&1.07\%\\
			$d_p/H$    &--&0.052&0.052&0.076&0.052&0.076\\
			$Ar$       &--&38171&38171&114528&38171&114528\\
			$Re_p$     &--&235.5&268.3&475.2&233.6&463.6\\
			$C_D$      &--&0.89&0.705&0.666&0.93&0.703\\
		\end{tabular}
		\caption{Parameters of the cases used for the present study according to \cite{2016_Santarelli}. The labels $BiDisp(Sm)$ and $BiDisp(La)$ denote the results for the bi-disperse case, $BiDisp$, where averaging has been restricted to small and large, respectively. Here, $N_p$ is the number of bubbles, $\alpha_b$ is the bulk void fraction, $d_p$ the bubble diameter, $Ar\equiv\left|\rho^G-\rho^L\right|gd_{p}^{3}/\rho^L\nu^{2}$ the Archimedes number. The values of $Re_p$, the bubble Reynolds number based on $d_p$ and the bubble to fluid relative velocity, as well as $C_D$ the drag coefficient, are obtained from the DNS.}
		\label{tab: DNSdatabase}
	\end{center}
\end{table}

As shown in figure \ref{fig: DNS schema}, the vertical flow takes place between two flat walls separated by the distance $H$, and the size of the computational domain is $L_x\times L_y\times L_z=4.41H\times H\times 2.21H$. Here, $x$ denotes the streamwise position coordinate, $y$ the wall-normal coordinate, and $z$ the spanwise coordinate, and the corresponding unit basis vectors are $\boldsymbol{e}_x,\boldsymbol{e}_y,\boldsymbol{e}_z$, respectively. The numerical grid employed has the same spacing $\Delta=H/232$ in all directions, resulting in $1024 \times 232 \times 512$ grid points in the $x$, $y$, and $z$ directions, respectively. A no-slip condition was applied at the walls, and periodic boundary conditions were applied in the $x$ and $z$ directions. The gravitational force acts in the direction $-\boldsymbol{e}_x$, and the bulk velocity $U_b$ was kept constant by instantaneously adjusting a volume force, equivalent to a pressure gradient, thus imposing a desired bulk Reynolds number $Re_b=U_bH/\nu$, where $\nu$ is the kinematic viscosity of the fluid. The DNS were all conducted with $Re_b=5263$. 

The data used in this work were obtained for three monodisperse cases (\textit{SmMany}, \textit{SmFew}, \textit{LaMany}) and one bi-disperse case labelled \textit{BiDisp}, of the same void fraction as \textit{SmMany} and \textit{LaMany} with half the void fraction consisting of smaller bubbles and the other half of larger bubbles. Additionally, a single-phase simulation labelled \textit{Unladen} was performed under the same conditions for comparison. Table \ref{tab: DNSdatabase} provides an overview of all cases with the corresponding labels.

\subsection{Data processing}

A standard way to analyze the multiscale properties of turbulence is to use the fluid velocity increments $\Delta \boldsymbol{u}'(\boldsymbol{x},\boldsymbol{r},t)\equiv \boldsymbol{u}'(\boldsymbol{x}+\boldsymbol{r},t)-\boldsymbol{u}'(\boldsymbol{x},t)$, where $\boldsymbol{u}'\equiv \boldsymbol{u}-\langle\boldsymbol{u}\rangle$ is the fluctuating fluid velocity, $\boldsymbol{r}$ is the separation vector, and $\left\langle \cdot \right\rangle$ denotes an ensemble average (estimated using appropriate space and time averages). The calculation of velocity increments in a bubble-laden flow is, however, delicate, since the phase boundaries can interrupt the fluid flow signal. To overcome this non-continuous velocity signal challenge, different methods have been used in the literature, such as smoothing the discontinuities by a Gauss function \citep{1991_Lance}; considering only intervals between bubbles where the velocity signal is continuous \citep{2010_Martinez,2013_Mendez,2011_Roghair}; and measuring the wake behind a rising swarm of bubbles, where there are no bubbles \citep{2010_Riboux}. Here, we use a method ideally suited for interface-resolved DNS of disperse flows proposed in our previous study \citep{2017_Ma}. In this method, the fluid velocity is recorded along grid lines in the spanwise direction whose wall-normal location lies within the centre region $0.474H<y<0.526H$ (this width corresponds to the smaller bubble diameter, see figure. \ref{fig: DNS schema}), which is a sufficiently thin region for these lines to be considered statistically equivalent. For each line, data was recorded whenever the entire line was free from bubbles, and we recorded $1,000,000$ instances of this for each case. With this method, we cannot compute $\Delta \boldsymbol{u}'(\boldsymbol{x},\boldsymbol{r},t)$ for arbitrary $\boldsymbol{r}$, but can compute $\Delta \boldsymbol{u}'(\boldsymbol{x},r_3\boldsymbol{e}_3,t)$, allowing us to perform an extensive investigation into the scale-dependent properties of bubble-laden turbulent flows.

\section{Reynolds number in bubble-laden turbulent flows}\label{sec: Taylor Re}

The range of scales in a turbulent flow is governed by its Reynolds number, and therefore before analyzing the properties of the bubble-laden turbulent channel flows at different scales we first consider the Reynolds numbers. As mentioned in \S\ref{sec: DNS database}, the bulk Reynolds number $Re_b=U_bH/\nu$ was kept fixed at $Re_b=5263$. However, since we are interested in the properties of the fluctuating component of the velocity field, it is more informative for our purposes to consider a Reynolds number based on the fluctuating velocity field. To that end we consider the Reynolds number $Re_H\equiv u^\ast H/\nu$, where $u^\ast\equiv \sqrt{(2/3)k_c}$, and $k_c$ is the turbulent kinetic energy (TKE) $k\equiv(1/2)\langle u_i'u_i'\rangle$ evaluated at the channel centre. 

\begin{figure}
	\centering
	\includegraphics[height=4cm]{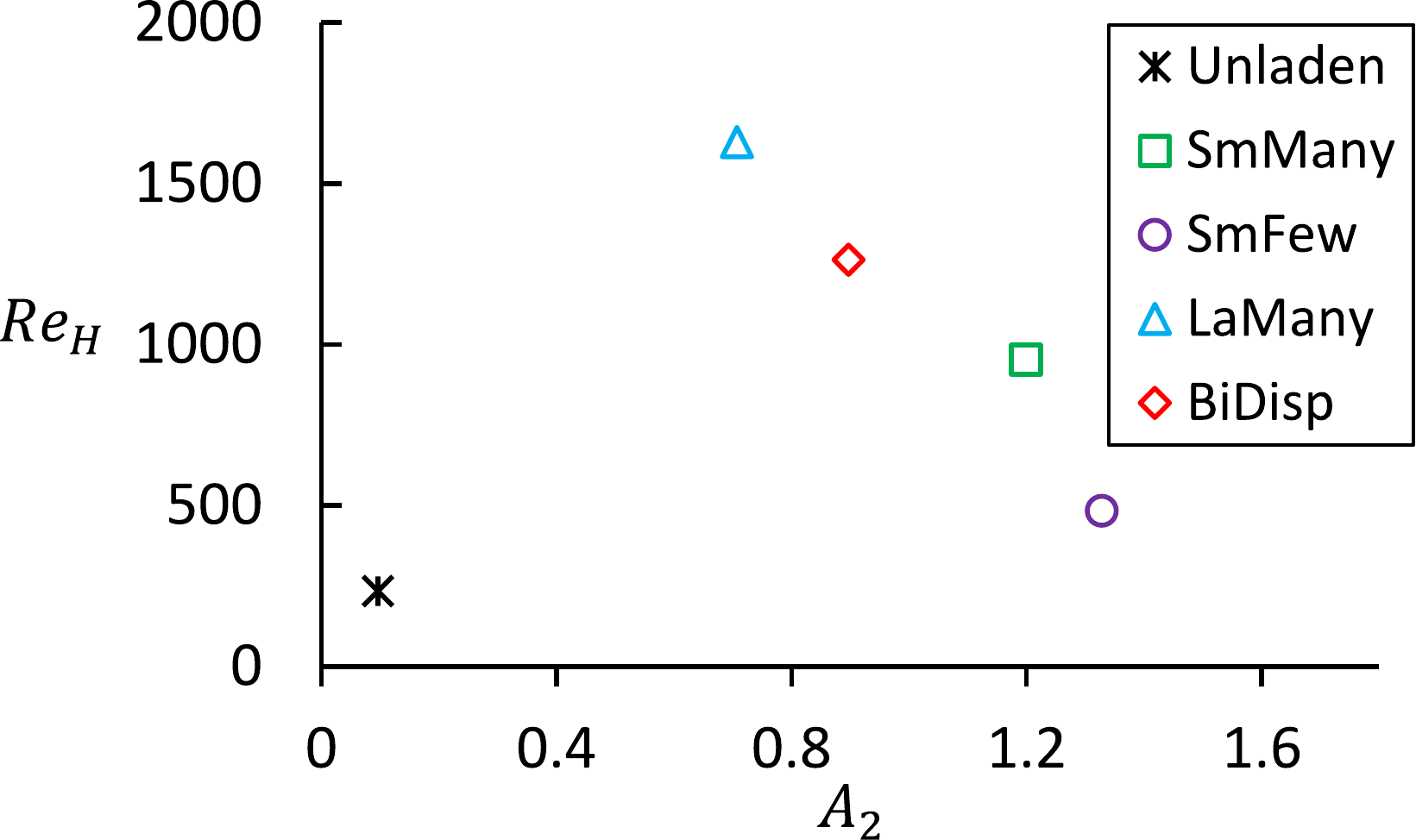}
	\caption{Reynolds number, $Re_H$ plotted versus the second anisotropy invariant.} \label{fig: Re_H}
\end{figure}

In figure \ref{fig: Re_H} we plot $Re_H$ versus $A_2$, where $A_2=a_{ji}a_{ij}$ is the second invariant of the Reynolds-stress anisotropy tensor (evaluated at the channel centre), $a_{ij}=\langle{u}'_i{u}'_j\rangle/k-(2/3)\delta_{ij}$, that quantifies the magnitude of the large-scale anisotropy in the flow. The results show that $Re_H$ varies significantly across the cases. The \textit{SmFew} case, $Re_H$ is only slightly larger than the unladen case as the bubble size $d_p$ is small and the bulk void fraction $\alpha_b$ is quite low. However, as $d_p$ and $\alpha_b$ are increased, $Re_H$ increases significantly, implying that as $d_p$ and $\alpha_b$ are increased (at least over the range we consider), the range of excited scales in the flow also increases, and hence the flow becomes increasingly multiscale. The increase cannot continue indefinitely, however, since when $d_p$ and $\alpha_b$ become sufficiently large, the problem becomes analogous to flow through a porous medium, for which the flow Reynolds number cannot be very large. This is also related to the fact that in such a regime, $H$ is no longer the relevant lengthscale in the flow Reynolds number, but rather the inter-bubble distance becomes the appropriate length scale.

In the considered range of bubble/channel parameters, figure \ref{fig: Re_H} suggests a approximately linear relation between $Re_H$ and $A_2$ for the bubble-laden cases, with larger $A_2$ corresponding to smaller $Re_H$. This behaviour may be understood in terms of the model for BIT from \cite{2020_Ma_a}, according to which
\begin{equation}
k_c=\frac{C_{\epsilon{2}}d_pS_{k}}{0.3(1-\alpha)C_D{u_r}}, \label{eq: algebraic k}
\end{equation} 
where $S_k$ is the interfacial term appearing in the TKE equation, $u_r\equiv \|\boldsymbol{u_r}\|$ with $\boldsymbol{u_r}$ the mean slip velocity between the bubble and the fluid at the channel centre, $C_{\epsilon{2}}=1.92$, and $\alpha$ is the void fraction in the channel centre. The algebraic expression in \eqref{eq: algebraic k} was derived based on the fact that in BIT-dominated flows the dissipation term is balanced by the interfacial term in the TKE transport equation at the channel centre. We refer the reader to \cite{2020_Ma_a} for the detailed derivation. 

\begin{figure}
	\centering
	\includegraphics[height=4cm]{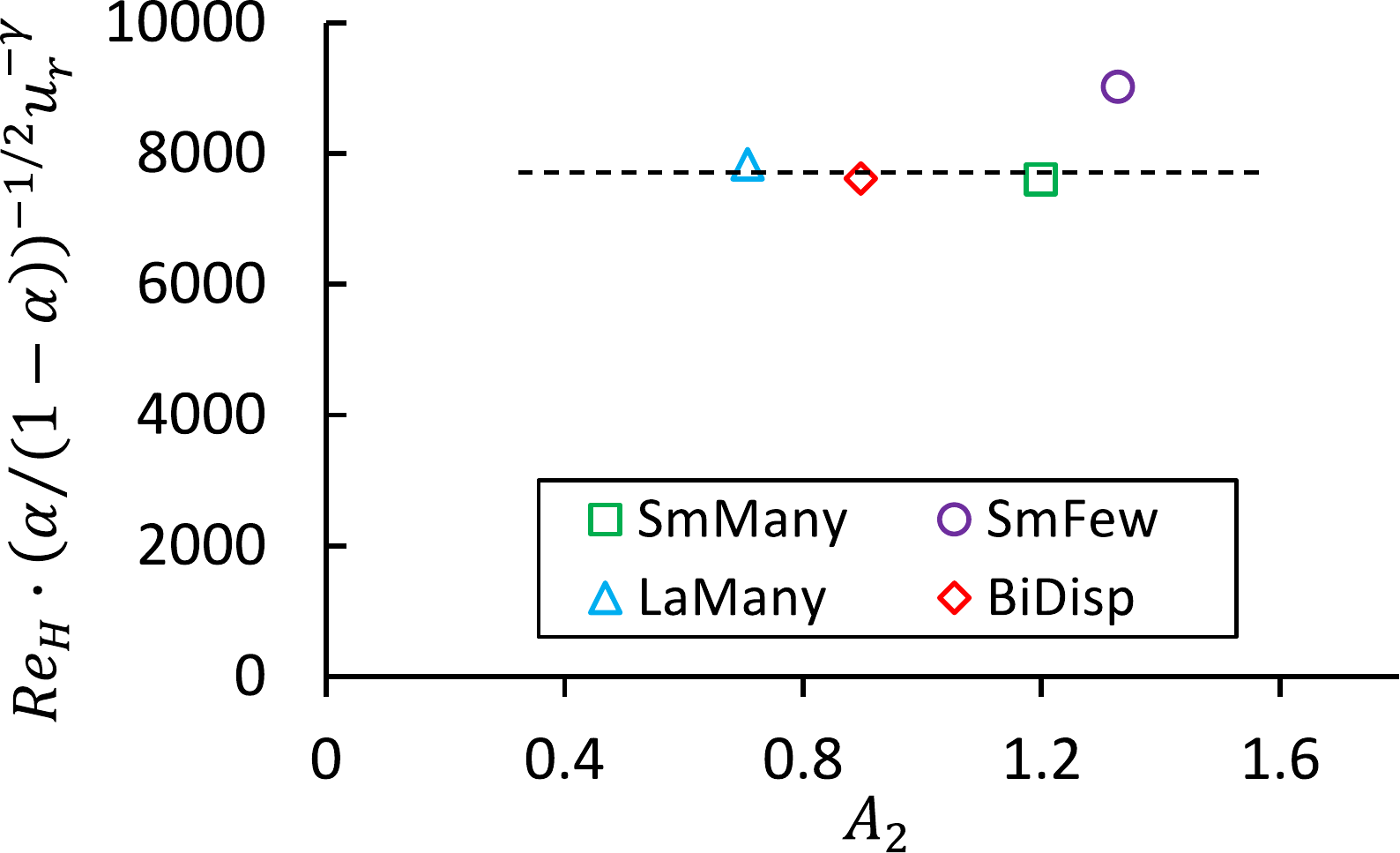}
	\caption{Reynolds number $Re_H$ normalized by the factor $\sqrt{\alpha/(1-\alpha)}u_r^\gamma$ plotted versus the second anisotropy invariant.} \label{fig: Re_H scaled}
\end{figure}

Based on (\ref{eq: algebraic k}) we can re-express the velocity scale $u^\ast$ by introducing the interfacial term $S_k$ adopted from \cite{2017_Ma}
\begin{equation}
S_{k}=\min(0.18 Re_{p}^{0.23},1)\boldsymbol{F}_{D}\boldsymbol{\cdot}\boldsymbol{u}_r \;,
\end{equation}
where $\boldsymbol{F}_{D}=(3/4d_p)C_D\alpha\|\boldsymbol{u}_r\|\boldsymbol{u}_r$ is the drag force on the bubbles at the channel centre. Using these results we obtain
\begin{equation}
u^\ast=\sqrt{\frac{2}{3}k}=\sqrt{\frac{5\nu C_{\epsilon2}\rho^L\min(0.18 Re_{p}^{0.23},1)}{3(1-\alpha)}  \alpha  u_r^2} \propto \sqrt{\alpha/(1-\alpha)}u_r^\gamma \;, \label{eq: velo-scale}
\end{equation} 
where $\gamma=1.12$ if $0.18 Re_{p}^{0.23}<1$, and $\gamma=1$ otherwise. This finally leads to 
\begin{equation}
Re_H=\frac{u^\ast H}{\nu} \propto \sqrt{\alpha/(1-\alpha)}u_r^\gamma \;. \label{eq: scaling factor}
\end{equation} 
In figure \ref{fig: Re_H scaled} we plot $Re_H$ normalized by the factor $\sqrt{\alpha/(1-\alpha)}u_r^\gamma $ in order to test the prediction in \eqref{eq: scaling factor}. The results show that the relation $Re_H \propto \sqrt{\alpha/(1-\alpha)}u_r^\gamma $ describes the data very well except for the \textit{SmFew} case. This is due to the fact that for this case the void fraction is low (0.29\%), so the influence of the background flow (ignored in \eqref{eq: algebraic k}) is not negligible. Further data across a wider range of the parameter space is needed to more comprehensively validate \eqref{eq: scaling factor} and to discover the extent of the range over which it holds.

\section{Multiscale anisotropy and second-order structure function}\label{sec: Anisotropy}

A systematic approach for analyzing the multiscale anisotropy of a turbulent flow is to use the irreducible representations of the SO(3) group, which consists of projecting the multipoint turbulent correlation functions onto the space of spherical harmonics \citep{1999_Arad,2001_Biferale,2005_Biferale}. However, such an analysis requires information on the full three dimensional flow field, something that is usually not obtainable from experiments. Furthermore, as discussed in \S\ref{sec: DNS database}, phase boundaries interrupt the flow field in multiphase flows, introducing further challenges in applying this method.

Due to these challenges, many investigations on anisotropy in turbulent flows focus directly on the structure function tensor (which are essentially moments of the velocity increments), comparing the longitudinal and transverse components in order to discern the level of anisotropy in the flow. In single-phase flows, the main focus was on scrutinizing the postulate of local isotropy and its implications \citep[][K41 for brevity]{1941_Kolmogorov_a}. In general, experiments and numerical simulations do not strictly confirm the convergence toward isotropy predicted within K41 theory as a function of the scale \citep{2000_Kurien,2002_Shen,1997_Dhruva}. In the context of the single-point Reynolds stresses, the Lumley triangle \citep{1977_Lumley} provided a powerful way to quantify and visualize anisotropy in the flow. It would be desirable to have something analogous to this for the structure functions, which would then provide a way to quantify and visualize anisotropy in the flow at different scales.

\subsection{Second-order structure function and its anisotropy}\label{subsec: Conventional}

Consider the second-order structure function
\begin{equation}
D_{ij}(\boldsymbol{r},t)\equiv\langle \Delta u'_i(\boldsymbol{r},t)\Delta u'_j(\boldsymbol{r},t) \rangle  \; . \label{eq: Dij}
\end{equation}
Hereafter we will suppress the time argument since we are focusing on statistically stationary flows. The Cartesian coordinate system chosen is depicted in figure \ref{fig: DNS schema}, and as discussed in \S\ref{sec: DNS database}, our DNS data only allow us to compute the velocity increments for separations in the spanwise direction, i.e. $\boldsymbol{r}=r_3\boldsymbol{e}_3$ (and $r\equiv\|\boldsymbol{r}\|=r_3$). In this case the longitudinal structure function is $D_{LL}(r_3)=D_{33}(r_3)$, and for an incompressible, isotropic flow we would have
\begin{equation}
D_{11}^{iso}=D_{22}^{iso}=D_{33}+\frac{r}{2}\frac{\partial}{\partial{r}}D_{33}\;; \;\;\;\; D_{ij}^{iso}=0 \; (\mathrm{for} \; i\neq j)\;. \label{eq: Dij-iso in coordi.}
\end{equation}
Figure \ref{fig: D11,D22,D33} displays the results for $D_{11},D_{22},D_{33}$ as a function of $r/H$ for all of the DNS cases. We also plot the isotropic form of the results based on (\ref{eq: Dij-iso in coordi.}) for the cases \textit{Unladen} (labelled \textit{Unladen iso}) and \textit{SmFew} (labelled \textit{SmFew iso}). The results show that in general the introduction of bubbles into the flow leads to strong enhancements of the fluctuations in all three directions of the flow. For $r/H\geq O(1)$, $D_{\gamma\gamma}\approx 2\langle u'_\gamma u'_\gamma\rangle$ as expected (no index summation is implied), and are consistent with the values obtained by \cite{2016_Santarelli} for the corresponding Reynolds normal stresses in the channel centre. Compared to the unladen case, the enhancement of the structure function level is in the sequence \textit{SmFew}, \textit{SmMany}, \textit{BiDisp} to \textit{LaMany}, which corresponds to increasing averaged bubble Reynolds number and/or gas void fraction. This holds for all three directions across all scales from around one channel width to the smallest dissipative scales, and shows that the bubbles modify fluctuations in the flow at scales both larger and smaller than the bubble length scale $d_p$. Moreover, the increase of the structure functions in the bubble-laden cases is more pronounced at the smaller scales than the larger scales. This behavior is in close agreement with the experimental results in \cite{2005_Rensen} which used Taylor's hypothesis to construct the results. We also note that although $D_{22}$ and $D_{33}$ are very similar for the \textit{Unladen} and the \textit{SmFew} case at the large scales, they are significantly different at the smaller scales, with the bubbles significantly enhancing the small scale fluctuations in the flow.

\begin{figure}	
	\begin{minipage}[b]{1.0\linewidth}
		\begin{minipage}[b]{0.5\linewidth}
			\centering
			\makebox[0.5em][l]{\raisebox{-\height}{(\textit{a})}}%
			\raisebox{-\height}{\includegraphics[height=3.6cm]{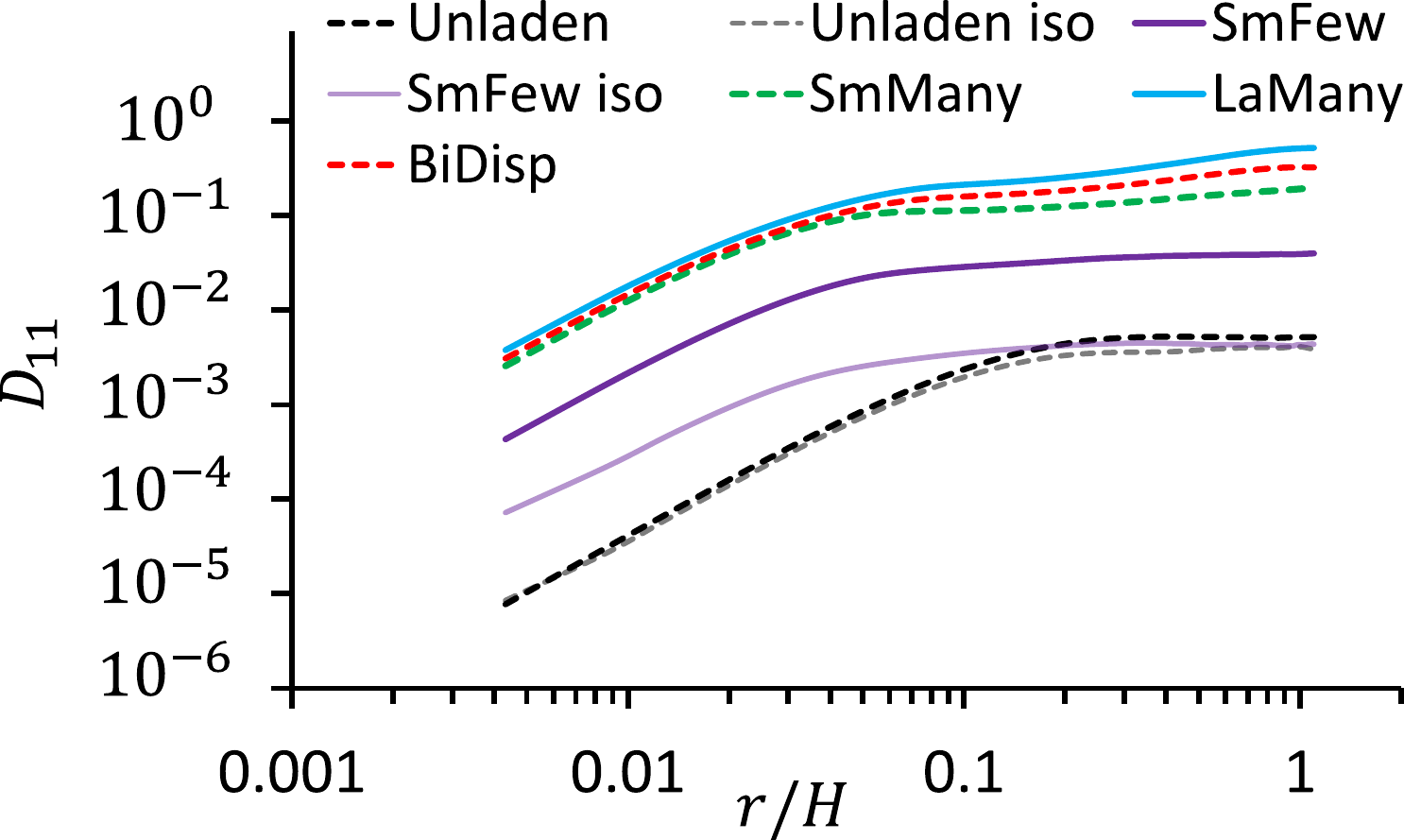}}
		\end{minipage}
		\begin{minipage}[b]{0.5\linewidth}
			\centering
			\makebox[0.5em][l]{\raisebox{-\height}{(\textit{b})}}%
			\raisebox{-\height}{\includegraphics[height=3.6cm]{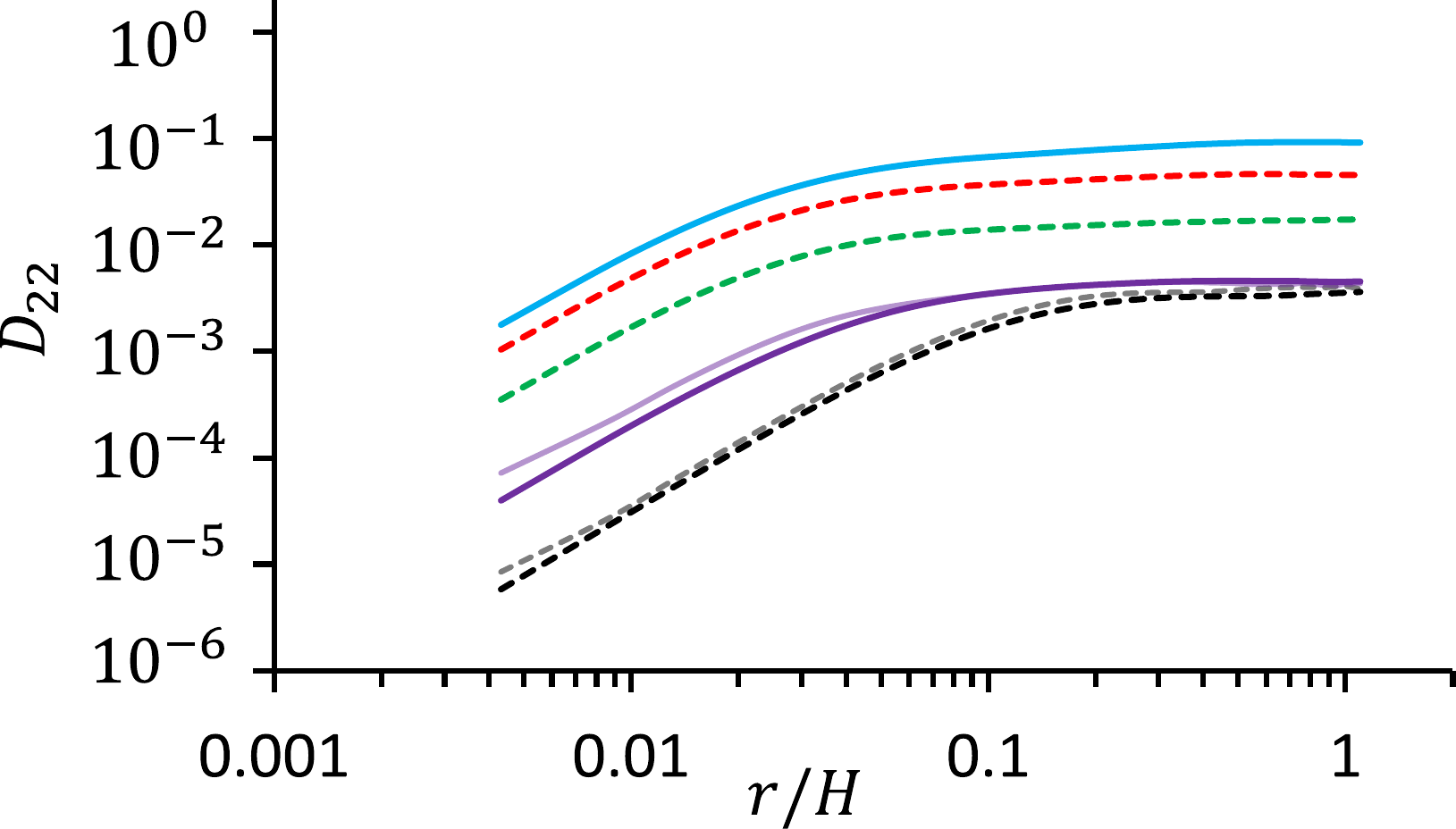}}
		\end{minipage}
	\end{minipage}	
	\begin{minipage}[b]{1.0\linewidth}
		\centering
		\vspace{3mm}
		\makebox[0.5em][l]{\raisebox{-\height}{(\textit{c})}}%
		\raisebox{-\height}{\includegraphics[height=3.6cm]{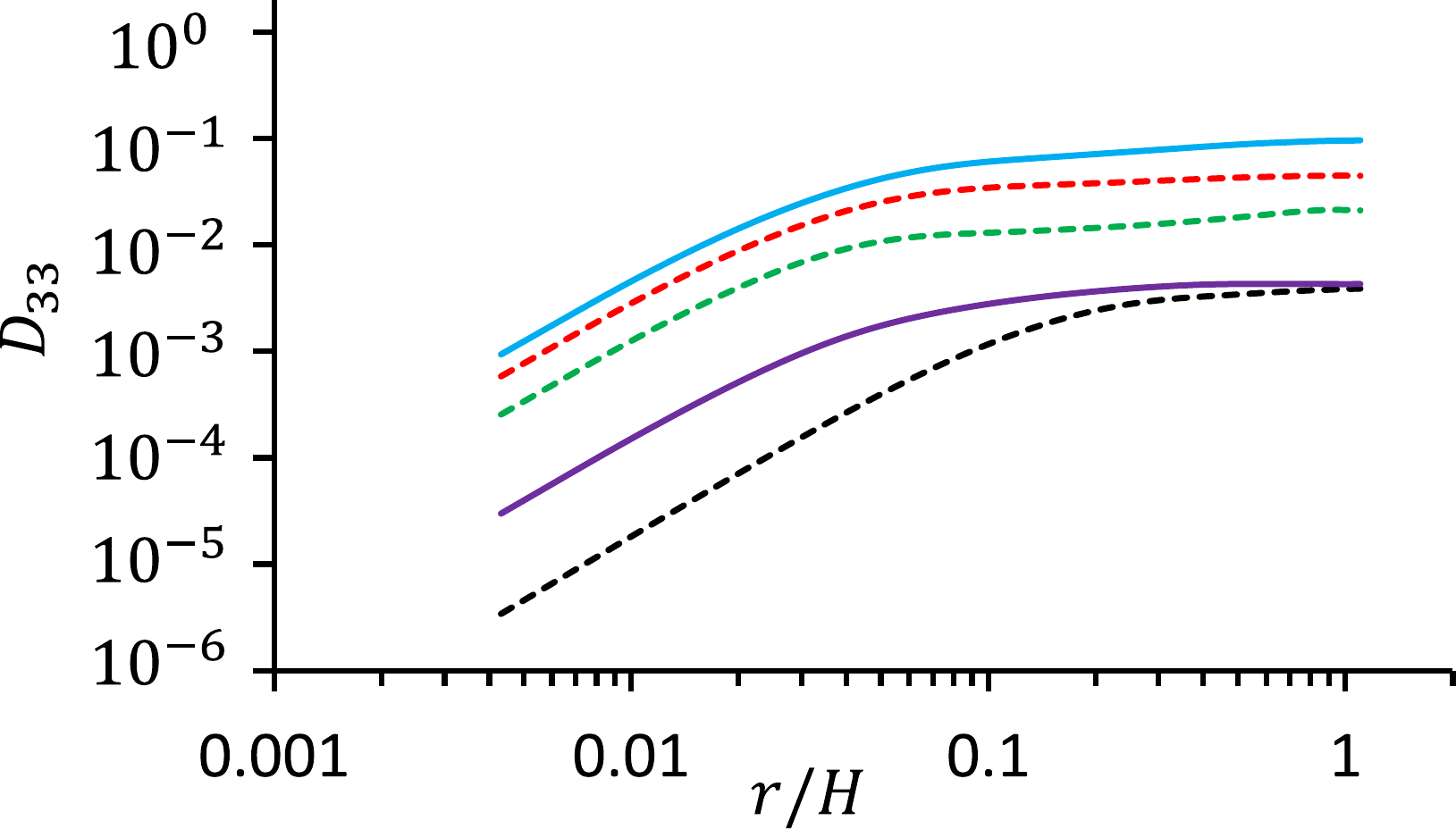}}
	\end{minipage}
	\caption{DNS results for the second-order transverse (\textit{a},\textit{b}) and longitudinal (\textit{c}) structure functions.} \label{fig: D11,D22,D33}
\end{figure}

For the unladen case, departures from $D_{11}= D_{11}^{iso}$ and $D_{22}= D_{22}^{iso}$ are not too strong, with stronger departures at the larger scales, consistent with the Reynolds stress behaviour in the channel centre $\langle u_1'u_1'\rangle/\langle u_2'u_2'\rangle\approx1.47$. As $r$ is decreased the flow becomes more isotropic, although as we shall show later, the small scales do not actually reach an isotropic state. In contrast, for the bubble-laden cases, there are strong departures from isotropy at all scales. For example, for the \textit{SmFew} case, $D_{11}$ shows significant deviations from $D_{11}^{iso}$ at all scales, and for $D_{22}$ the deviations from the isotropic form $D_{22}^{iso}$ actually become stronger as one goes to scales smaller than the bubble length scale ($d_p\approx 0.05H$ for this case). The concept of return to isotropy can therefore be strongly violated for bubbly turbulent flows. This is not surprising however, since the mean velocity of the bubbles is unidirectional, and the fact that $d_p\ll H$ means that the bubbles can directly inject fluctuations into the flow at the small scales, in contrast to the unladen case where the fluctuations are injected into the flow at large scales due to the mean shearing of the flow.

\begin{figure}	
	\begin{minipage}[b]{1.0\linewidth}
		\begin{minipage}[b]{0.5\linewidth}
			\centering
			\makebox[0.5em][l]{\raisebox{-\height}{(\textit{a})}}%
			\raisebox{-\height}{\includegraphics[height=3.6cm]{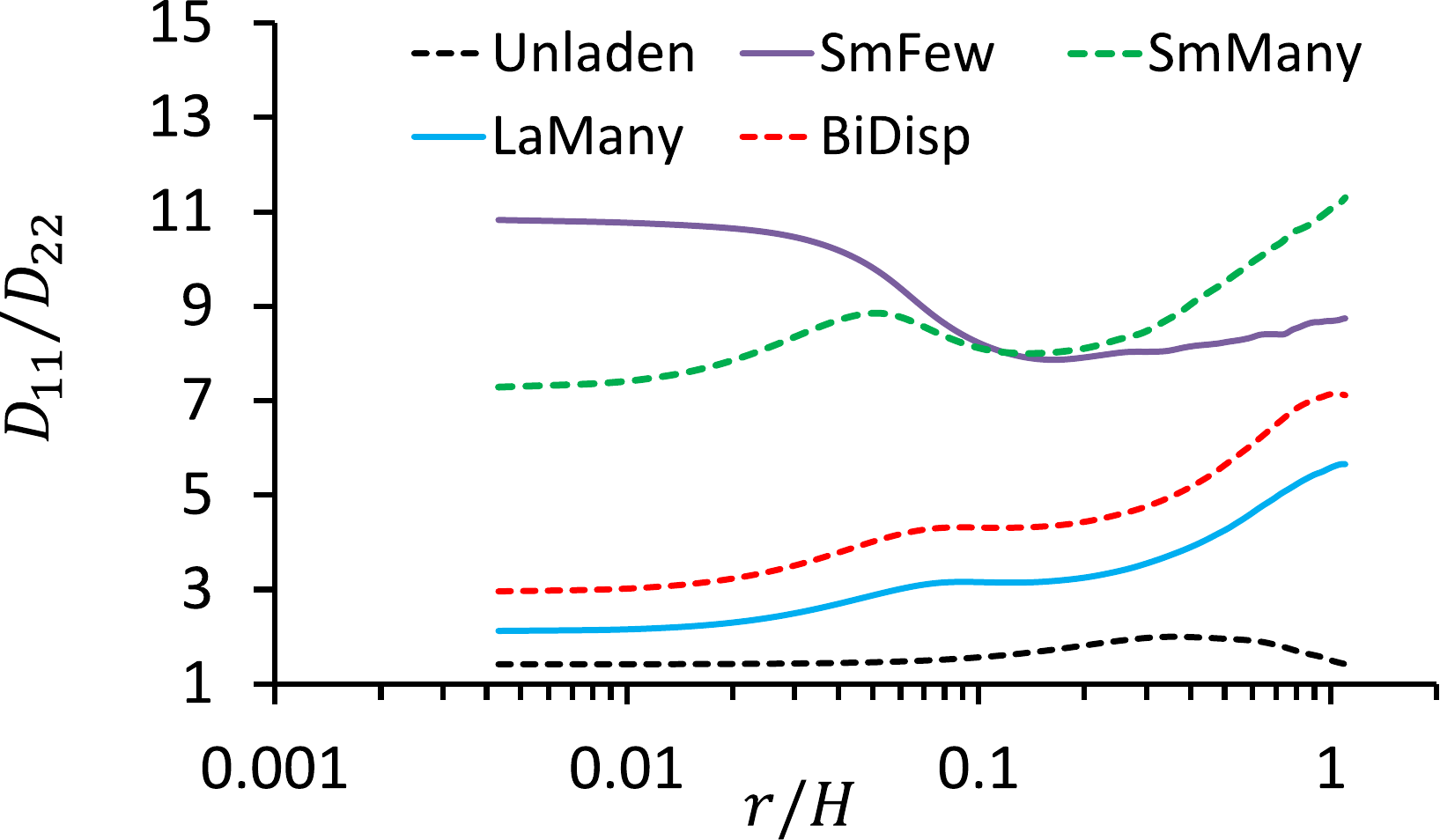}}
		\end{minipage}
		\begin{minipage}[b]{0.5\linewidth}
			\centering
			\makebox[0.5em][l]{\raisebox{-\height}{(\textit{b})}}%
			\raisebox{-\height}{\includegraphics[height=3.6cm]{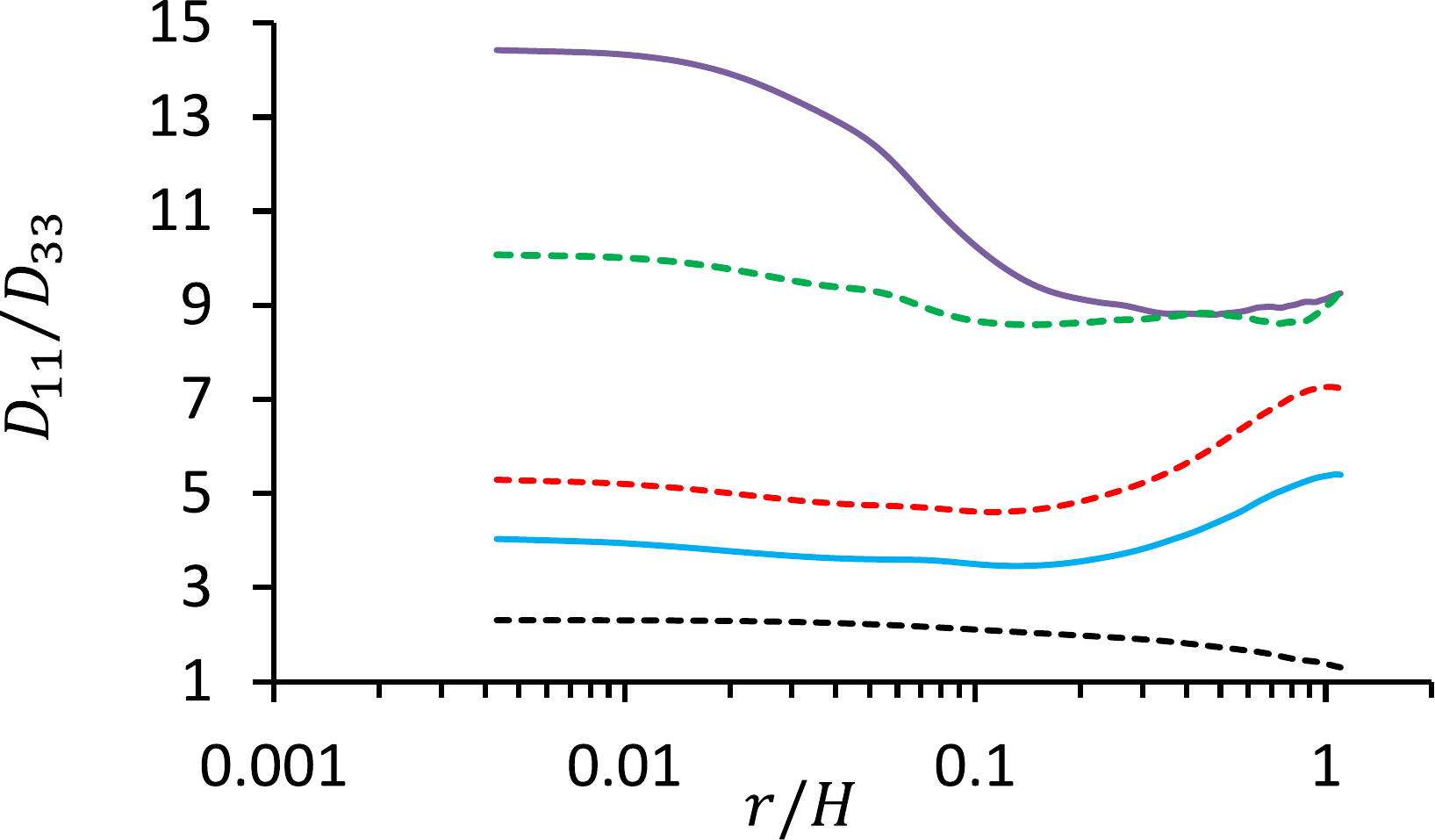}}
		\end{minipage}
	\end{minipage}
	\begin{minipage}[b]{1.0\linewidth}
		\centering
		\vspace{3mm}
		\makebox[0.5em][l]{\raisebox{-\height}{(\textit{c})}}%
		\raisebox{-\height}{\includegraphics[height=3.6cm]{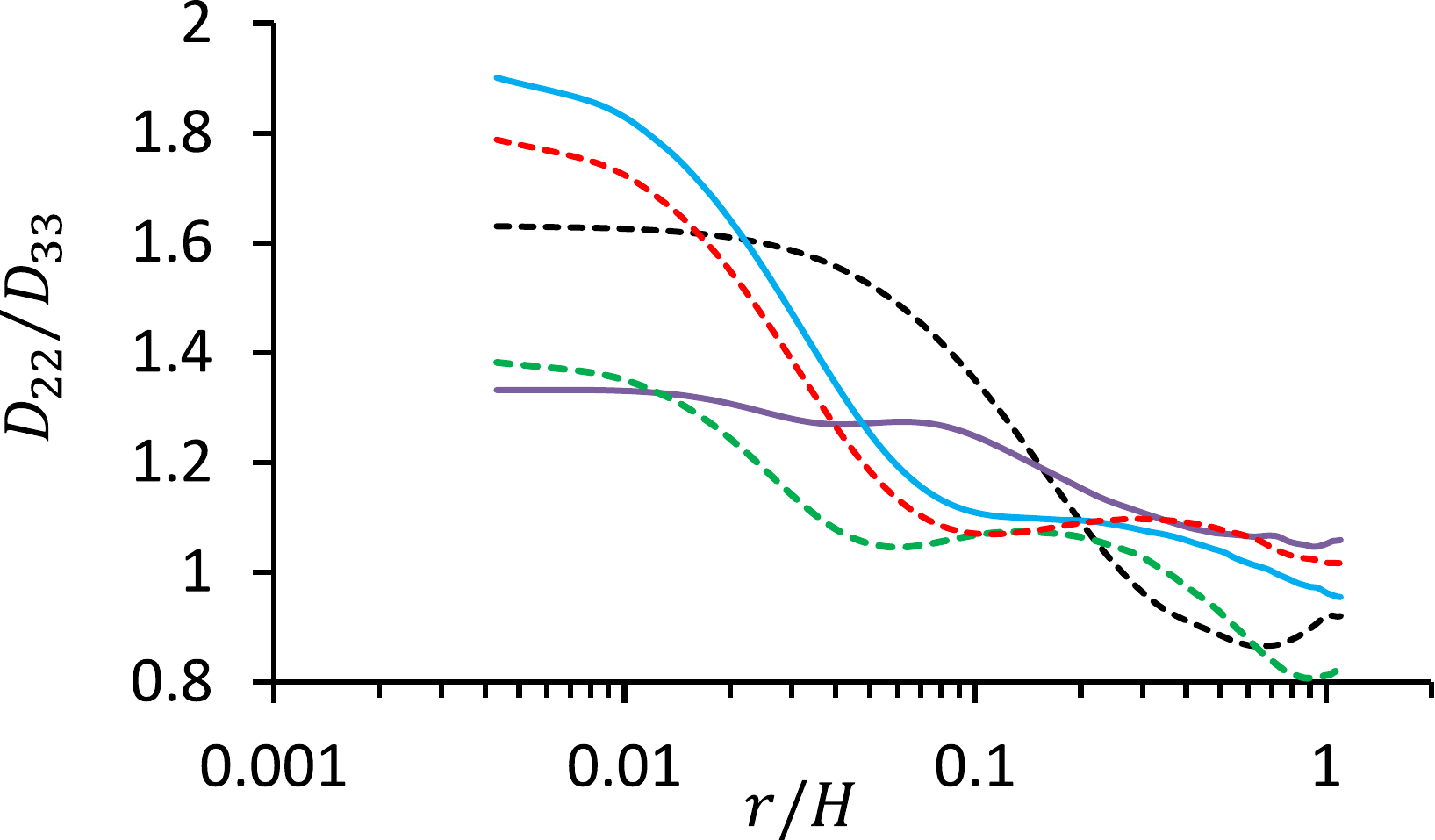}}
	\end{minipage}
	\caption{Ratio of $D_{TT}/D_{TT}$ (a) and $D_{TT}/D_{LL}$ (b,c) versus separation distance for all cases considered} \label{fig: Ratio of SFs}	
\end{figure}

Figure \ref{fig: Ratio of SFs} shows the ratios of the different structure function components in order to see more clearly the anisotropy of the flow. For an isotropic system we would have $D_{11}/D_{22}=1$ at all scales, and for the \textit{unladen} case $D_{11}/D_{22}$ is quite close to 1, especially at smaller scales, while for the bubble-laden cases $D_{11}/D_{22}$ deviates strongly from 1, reaching values $O(10)$. Except for the \textit{SmFew} case, the behaviour of $D_{11}/D_{22}$ in the bubble-laden cases monotonically decreases with decreasing $r$, indicating a tendency towards return to isotropy, although the isotropic state is far from achieved. In isotropic turbulence, $D_{11}/D_{33}\rightarrow1$ and $D_{22}/D_{33}\rightarrow1$ for $r\geq O(H)$, while $D_{11}/D_{33}\rightarrow 2$ and $D_{22}/D_{33}\rightarrow2$ for $r/H\to 0$ \citep{2000_Pope}. While deviations from these are not too strong for the \textit{unladen} case, strong departures are observed for the bubble-laden cases at all scales. The departures are strongest for the ratios involving $D_{11}$, which is to be expected since this is the direction of the mean trajectory of the bubbles. Moreover, for each of the ratios plotted in figure \ref{fig: Ratio of SFs}, the bubble-laden cases reveal a bump at $r=O(d_p)$, indicating the injection of anisotropy into the flow due to the bubbles and their anisotropic motion in the flow due to the buoyancy force acting on them.

While comparisons of the ratios of $D_{11}, D_{22}, D_{33}$ is a standard way to analyse the multiscale anisotropy in the flow \citep{1999_van_de_Water,2017_Carter}, this method provides limited quantitative insight into the degree of anisotropy. For example, while the behaviour of these ratios is well-known for an isotropic flow in the limits $r/H\to 0$, $r/H\geq O(1)$, their behaviour for intermediate $r/H$ is not known and cannot be determined simply by the condition of isotropy. It is therefore desirable to provide a simple measure of anisotropy based on $D_{ij}$ that applies at all scales, and also helps to visualize the anisotropic behaviour. In the context of the Reynolds stress tensor, this can be achieved using either the Lumley triangle \citep{1977_Lumley} or else the barycentric map \citep{2007_Banerjee}. As describe in the next subsection, however, these methods cannot be directly applied to quantify multiscale anisotropy associated with $D_{ij}$, and modification is required.

\subsection{Quantifying and visualizing scale-dependent anisotropy}\label{subsec: A new tool}

 Characterizing the scale-dependent anisotropy associated with $D_{ij}$ is considerably more involved than that based on the Reynolds stress $\langle u_i u_j\rangle$. This is because the relationship between componentiality and isotropy breaks down at sub-integral scales. For example, in the case of the Reynolds stress tensor, its components in the isotropic state are $\langle u'_i u_j'\rangle^{iso}\equiv\langle u'_m u_m'\rangle\delta_{ij}/3$, according to which the components in all three Cartesian directions are the same (a \textquotedblleft three-component flow\textquotedblright). In the case of $D_{ij}$, its isotropic state is (assuming the flow is incompressible)
\begin{equation}
D_{ij}^{iso}=D_{LL}\delta_{ij}+\frac{r}{2}\bigg(\delta_{ij}-\frac{r_ir_j}{r^2}\bigg)\frac{\partial}{\partial{r}}D_{LL} \;.\label{eq: new iso Dij}
\end{equation}
At the large scales $(\partial/\partial r)D_{LL}=0$, and we have $D_{ij}^{iso}=D_{LL}\delta_{ij}=\langle u'_m u_m'\rangle\delta_{ij}/3$, and hence the isotropic state corresponds to a three-component flow. On the other hand, in the limit $r\to0$ we have $D_{LL}=\langle\epsilon\rangle r^2/15\nu$ \citep{2000_Pope} and 
\begin{equation}
D_{ij}^{iso}=2 D_{LL}\delta_{ij}-\frac{r_ir_j}{r^2}D_{LL} \;.
\end{equation}
In this case the isotropic state does not correspond to a three-component flow, but rather the components transverse to $\boldsymbol{r}$ are twice as large as those parallel to $\boldsymbol{r}$. Hence, in general there is no correspondence between the componentiality of the flow and isotropy. For this reason, measures such as $D_{ij}-(D_{mm}\delta_{ij}/3)$, which have previously been used to define scale-dependent anisotropy \citep[e.g.][]{2018_Brugger}, do not in fact quantify isotropy, but rather only quantify deviations from the three-component state.

In view of these considerations, anisotropy must be quantified by the deviation of $D_{ij}$ from $D_{ij}^{iso}$, rather than from $D_{mm}\delta_{ij}/3$. To this end, we define two normalized tensors $\boldsymbol{A}^{iso}$ and $\boldsymbol{A}$ with components
\begin{equation}
A_{ij}^{iso}(\boldsymbol{r})\equiv\frac{D_{ij}^{iso}}{D_{kk}^{iso}} \;, \label{eq: Aij-iso}
\end{equation}
and
\begin{equation}
A_{ij}(\boldsymbol{r})\equiv\frac{D_{ij}}{D_{kk}} \;, \label{eq: Aij}
\end{equation}
respectively. Then, following \cite{2007_Banerjee}, we may express these components in terms of basis matrices
\begin{equation}
\widehat{A}_{ij}^{iso}=\mathcal{I}_{1c}B_{1c}+\mathcal{I}_{2c}B_{2c}+\mathcal{I}_{3c}B_{3c} \;, \label{eq: orth-Aij-iso}
\end{equation}
\begin{equation}
\widehat{A}_{ij}=\mathcal{J}_{1c}B_{1c}+\mathcal{J}_{2c}B_{2c}+\mathcal{J}_{3c}B_{3c} \; , \label{eq: orth-Aij}
\end{equation}
where the circumflex $\widehat{(\cdot)}$ denotes the tensor components expressed in the tensor principal axes, with $\widehat{A}_{ij}^{iso}=\mathrm{diag}(\lambda_{1}, \lambda_{2}, \lambda_{3})$, where $\lambda_{1}\geq\lambda_{2}\geq\lambda_{3}$ are the ordered eigenvalues of $\boldsymbol{A}^{iso}$, and $\widehat{A}_{ij}=\mathrm{diag}(\mu_1, \mu_2, \mu_3)$, where $\mu_1\geq\mu_2\geq\mu_3$ are the ordered eigenvalues of $\boldsymbol{A}$.  

The basis matrices $B_{1c}, B_{2c}, B_{3c}$ represent the one-component, two-component, and three-component limiting states, and are given by $B_{1c}=\mathrm{diag}(1,0,0)$, $B_{2c}=\mathrm{diag}(1/2,1/2,0)$, and $B_{3c}=\mathrm{diag}(1/3,1/3,1/3)$, respectively. The coefficients in (\ref{eq: orth-Aij-iso}) and (\ref{eq: orth-Aij}) then measure how close $\boldsymbol{A}^{iso}$ and $\boldsymbol{A}$ are to any of the three limiting states, and can be determined from the eigenvalues of the associated tensors
\begin{equation}
\mathcal{I}_{1c}=(\lambda_{1}-\lambda_{2})/\lambda_1\; ;\; \mathcal{I}_{2c}=(\lambda_{2}-\lambda_{3})/\lambda_1 \; ;\; \mathcal{I}_{3c}=\lambda_{3}/\lambda_1 \;, \label{eq: C_123c}
\end{equation}
\begin{equation}
\mathcal{J}_{1c}=(\mu_{1}-\mu_{2})/\mu_1\; ;\; \mathcal{J}_{2c}=(\mu_{2}-\mu_{3})/\mu_1 \; ;\; \mathcal{J}_{3c}=\mu_{3}/\mu_1 \;. \label{eq: C_123d}
\end{equation}
To visualize the componentiality of the flow implied by these coefficients, a barycentric map \citep{2007_Banerjee} can be defined as 
\begin{equation}
x_{BAM}=\mathcal{I}_{1c}x_{1c}+\mathcal{I}_{2c}x_{2c}+\mathcal{I}_{3c}x_{3c} \; , \label{eq: x-BAM}
\end{equation}
\begin{equation}
y_{BAM}=\mathcal{I}_{1c}y_{1c}+\mathcal{I}_{2c}y_{2c}+\mathcal{I}_{3c}y_{3c} \; , \label{eq: y-BAM}
\end{equation}
and similarly for the $\mathcal{J}$ components. Here, $(x_{1c}, y_{1c})=(1,0)$, $(x_{2c}, y_{2c})=(0,0)$, and $(x_{3c}, y_{3c})=(1/2,\sqrt{3}/2)$ are the three corner points corresponding to the limiting states of componentality. They can be chosen arbitrarily, but are commonly set to the corner points of an equilateral triangle, as this aids the interpretation of the visualization. We note that in the traditional Lumley triangle, it is the fact the trace of the Reynolds stress anisotropy tensor, $a_{ii}$, is zero that allows its properties to be represented in a two-dimensional triangle using two independent eigenvalues of $a_{ij}$. This restriction does not apply to the Barycentric map approach, which is why we are able to represent the properties of $\boldsymbol{A}^{iso}$ and $\boldsymbol{A}$ in a two-dimensional map, even though the trace of these tensors are not zero.

Provided that $(\partial/\partial r)D_{LL}\geq 0$ (true for a homogeneous flow), then the projections of $D_{ij}^{iso}$ in the plane orthogonal to $\boldsymbol{r}$ are greater than or equal to $D_{LL}$. As a result, $\boldsymbol{A}^{iso}$ is then confined to the left side of the triangle, corresponding to a state of axisymmetric contraction. Moreover, as $r$ is decreased, $\boldsymbol{A}^{iso}$ moves monotonically away from the three-component state towards the two-component state.  On the other hand, $\boldsymbol{A}$ may exist anywhere within the triangle.

Following this approach, $\boldsymbol{A}^{iso}$ and $\boldsymbol{A}$ can be mapped to a location $(x_{BAM}(\boldsymbol{r}), y_{BAM}(\boldsymbol{r}))$ in the barycentric map, and the linear distance between the coordinates corresponding to these two tensors then gives a measure of the anisotropy at that scale. In particular, the anisotropy at a given scale is determined by
\begin{equation}
\begin{split}
C_{ani}(\boldsymbol{r})&\equiv\|\boldsymbol{A}(\boldsymbol{r})-\boldsymbol{A}^{iso}(\boldsymbol{r})\|  \\
&=\sqrt{\big(\mathcal{J}_{1c}(\boldsymbol{r})-\mathcal{I}_{1c}(\boldsymbol{r})\big)^2+\big(\mathcal{J}_{2c}(\boldsymbol{r})-\mathcal{I}_{2c}(\boldsymbol{r})\big)^2+\big(\mathcal{J}_{3c}(\boldsymbol{r})-\mathcal{I}_{3c}(\boldsymbol{r})\big)^2} \;. \label{eq: C-ani}
\end{split}
\end{equation}
where we have used \eqref{eq: orth-Aij} and \eqref{eq: orth-Aij-iso}. For an isotropic flow, $\boldsymbol{A}=\boldsymbol{A}^{iso}$ and $C_{ani}= 0$. At the large scales of an isotropic flow, $\mathcal{I}_{1c}=\mathcal{I}_{2c}=0$ and $\mathcal{I}_{3c}=1$, so that the coordinates for $\boldsymbol{A}^{iso}$ are $(x_{BAM}, y_{BAM})=(1/2,\sqrt{3}/2)$. Therefore, at the large scales $C_{ani}(\boldsymbol{r})$ recovers the property that anisotropy is related to distance from the top of the triangle which corresponds to the three-component state, just as for the single-point Reynolds stress tensor \citep{2007_Banerjee}. By contrast, for an arbitrary scale in the flow, anisotropy is related to the distance in the barycentric triangle from a point along the left side of the triangle, the location of this point representing $\boldsymbol{A}^{iso}$. This is illustrated in figure \ref{fig: BAM}.

\subsection{Application of the new method}\label{subsec: New tool}

\begin{figure}
	\centering
	\includegraphics[height=5.9cm]{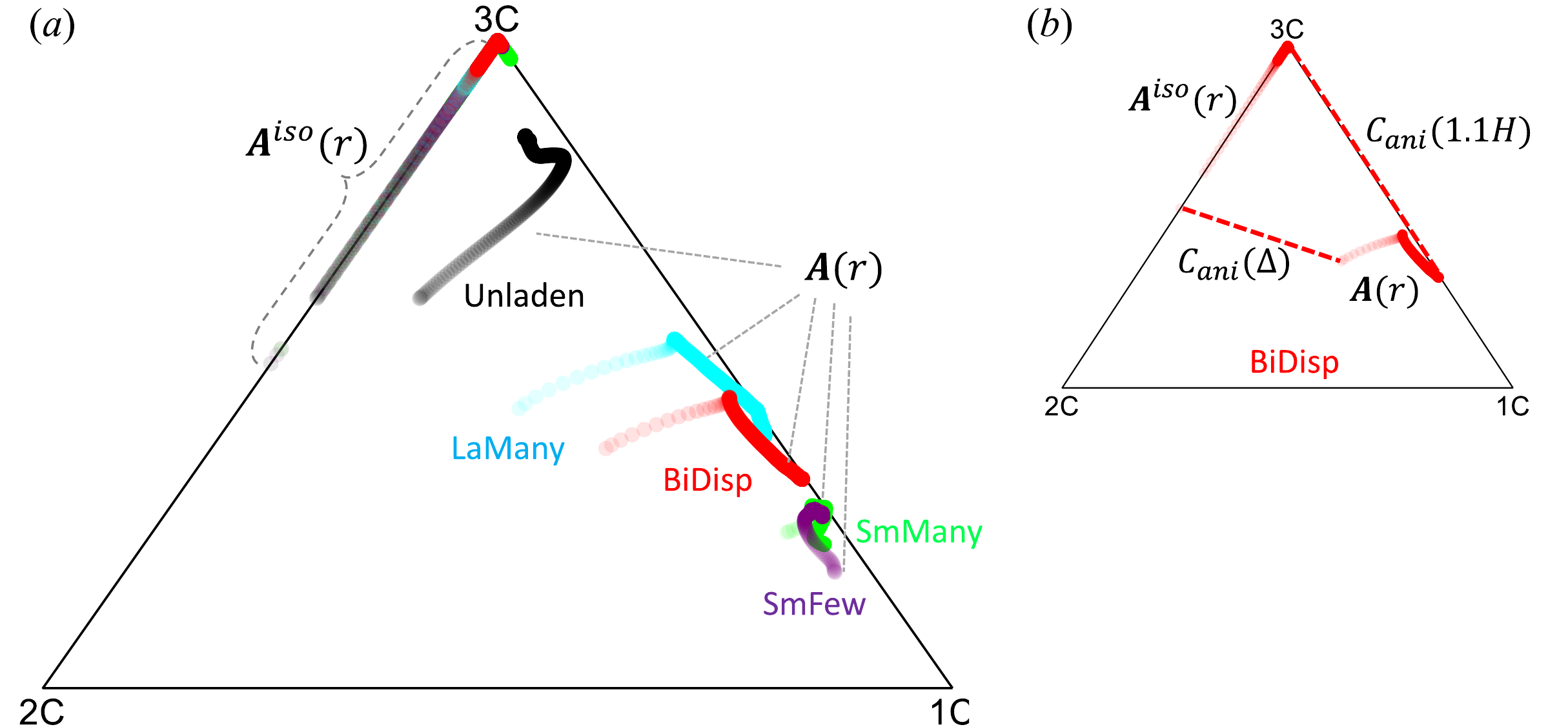}
	\caption{Representation of $\boldsymbol{A}$(\ref{eq: Aij}) and its isotropic form $\boldsymbol{A}^{iso}$ (\ref{eq: Aij-iso}) in barycentric map (\textit{a}) for all cases considered. The transparency of the color reflects the separation from small to large. $C_{ani}$ (\ref{eq: C-ani}) is represented by the red dashed lines (\textit{b}) for the case \textit{BiDisp}, highlighting e.g. $C_{ani}(\Delta)$ for the smallest scale and $C_{ani}(1.1H)$ for the largest scale.} \label{fig: BAM}
\end{figure}

We now turn to apply the new method described in \S\,\ref{subsec: A new tool} to our DNS results. The trajectories for $\boldsymbol{A}^{iso}$ in figure \ref{fig: BAM}(\textit{a}) all lie on the left side of the triangle, corresponding to the state of axisymmetric contraction, associated with the fact that $D_{11}^{iso}=D_{22}^{iso}\geq D_{33}$. The trajectories for $\boldsymbol{A}^{iso}$ at large $r$ are near to the three-component upper corner of the triangle, however, for the largest $r/H$ for which we have data, $(\partial/\partial r)D_{LL}\neq0$ and so the exact three component state $\boldsymbol{A}^{iso}=\mathbf{I}/3$ is not observed. As $r$ is decreased, the trajectory for $\boldsymbol{A}^{iso}$ moves down the left side of the triangle, towards the two-component corner, reflecting the fact that for $r/H\to 0$, $D_{11}^{iso}=D_{22}^{iso}=2D_{33}^{iso}$.

The results in figure \ref{fig: BAM}(\textit{a}) for $\boldsymbol{A}$ show that the bubble cases start closer to the one-component corner of the triangle than the unladen case. This reflects the fact that the fluctuations in the gravity direction are much stronger due to the buoyancy forces acting on the bubbles. The trajectories of $\boldsymbol{A}$ in the triangle are highly nonlinear, and show a tendency to migrate towards the axisymmetric contraction side of the triangle as $r$ is decreased, consistent with an approach to isotropy as $r$ is decreased, although never obtaining an isotropic state. The exception to this is the \textit{SmFew} case for which the trajectory of $\boldsymbol{A}$ actually approaches the one-component corner of the triangle as $r$ is decreased. This surprising behaviour is consistent with that observed in figure \ref{fig: Ratio of SFs}.

A quantitative measure of the scale-wise anisotropy is provided by $C_{ani}$, introduced in \S\,\ref{subsec: A new tool}. This quantity provides a linear measure of anisotropy, and is zero for an isotropic flow. For illustration, in figure \ref{fig: BAM}(\textit{b}) we show the trajectories of $\boldsymbol{A}$ and $\boldsymbol{A}^{iso}$ for the \textit{BiDisp} case and join the starting and ending points of these trajectories with dashed lines. The length of the line connecting the starting points represents $C_{ani}(r=1.1H)$, while the length of the line connecting the ending points represents $C_{ani}(r=\Delta)$, where $\Delta$ is the grid spacing in the spanwise direction. The plot shows that $C_{ani}(1.1H)>C_{ani}(\Delta)$, such that the flow is more isotropic at the smaller scales in \textit{BiDisp} case.

\begin{figure}
	\begin{minipage}[b]{1.0\linewidth}
		\begin{minipage}[b]{0.5\linewidth}
			\centering
			\makebox[0.5em][l]{\raisebox{-\height}{(\textit{a})}}%
			\raisebox{-\height}{\includegraphics[height=3.7cm]{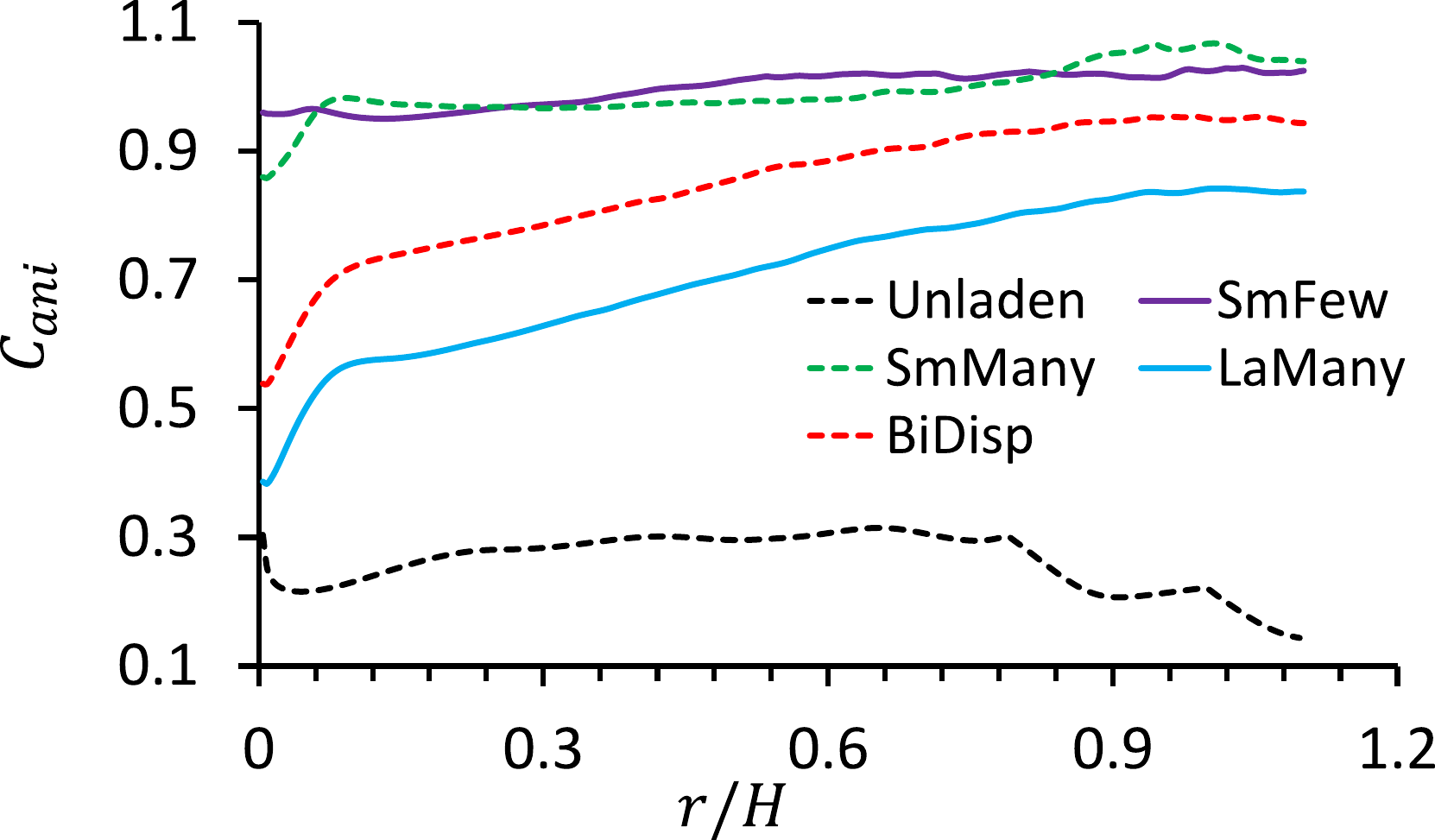}}
		\end{minipage}
		\begin{minipage}[b]{0.5\linewidth}
			\centering
			\makebox[0.5em][l]{\raisebox{-\height}{(\textit{b})}}%
			\raisebox{-\height}{\includegraphics[height=3.7cm]{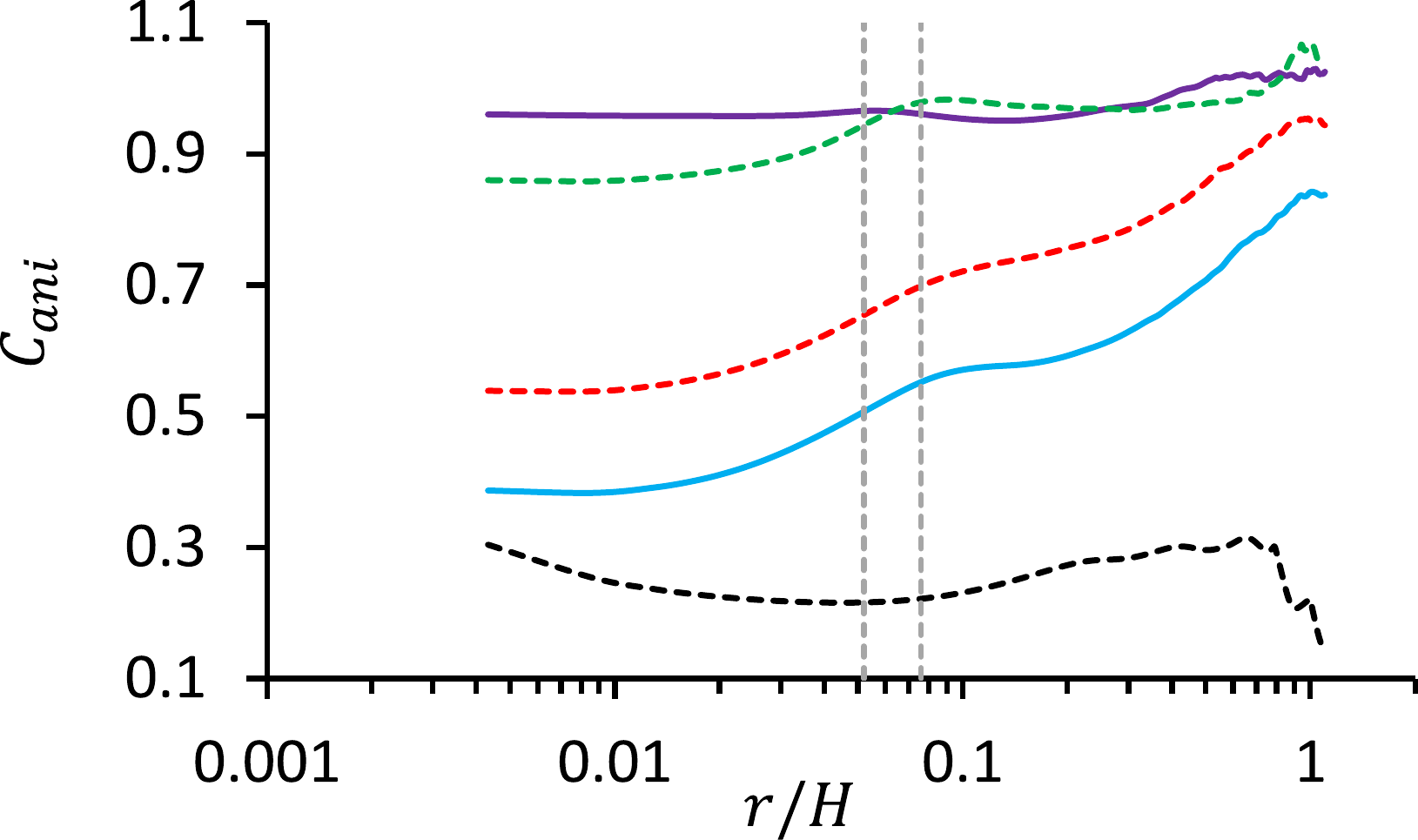}}
		\end{minipage}
	\end{minipage}
	\caption{Anisotropy measures $C_{ani}$ (\textit{a}) linear plot, (\textit{b}) the same data with semi-log plot. The two vertical dashed lines in (\textit{b}) show $r=d_p$ for smaller and larger bubbles, respectively.} \label{fig: C_ani}
\end{figure}

In figure \ref{fig: C_ani} we show the results for $C_{ani}$ as a function of $r/H$ for all of the cases. In general, $C_{ani}$ monotonically decreases with decreasing $r$, consistent with the return towards isotropy prediction. However, for these cases isotropy is never fully recovered, with anisotropy persisting into the dissipative range scales. Such behaviour for unladen turbulent flows has also been observed experimentally, including at much higher Reynolds numbers, as seen in studies by \cite{2000_Kurien}, \cite{2002_Antonia}, and \cite{2017_Carter}. Furthermore, we note that for $r\leq 0.02H$ (most clearly seen in the semi-log plot of figure \ref{fig: C_ani}\textit{b}) the return to isotropy is interrupted for the unladen case, and $C_{ani}$ actually becomes larger as $r$ is further decreased, implying increasing anisotropy at these scales. Similar behaviour has also observed in both experimental \citep{2017_Carter} and numerical studies \citep{1991_Meneveau,2007_Bos}, where it was suggested to be due to anisotropic intermittency at the dissipative scales.

The results in figure \ref{fig: C_ani} show that the departure from isotropy is in general much more pronounced for the bubble-laden cases than for the unladen case. This is again due to the fact that the bubbles introduce a significant source of momentum into the flow in the direction parallel to gravity due to the buoyancy force they experience. For the unladen case, the anisotropy is quite weak in the channel centre because the mean-shear that generates the anisotropy is weak in that region of the flow. We also notice from figure \ref{fig: C_ani}(\textit{a}) that the level of anisotropy decreases with increasing bubble Reynolds number, $Re_p$. In particular, the anisotropy across the scales generally decreases in the sequence \textit{SmMany (SmFew)}, \textit{BiDisp} to \textit{LaMany}, which corresponds to increasing $Re_p$. This behaviour is likely due to both the $Re_p$-dependent structure of the wakes produced by the bubbles \citep[see the supplementary materials to][]{2016_Santarelli} and also the path of the rising bubbles which becomes more chaotic and less uni-directional as $Re_p$ increases \citep{2010_Horowitz,2012_Ern}. It is also interesting to note that cases \textit{SmFew} and \textit{SmMany} which have similar $Re_p$, but different $\alpha_b$, have a similar level of anisotropy. This suggests that $Re_p$ plays more of a role than $\alpha_b$ in determining the contribution of the bubbles to the flow anisotropy, at least over some regimes of $\alpha_b$ (of course it cannot be true in general since, for example, when $\alpha_b=0$ the bubbles have no effect on the flow).

Another observation prompted by figure \ref{fig: C_ani} is that the shape of the $C_{ani}$ curve is remarkably similar for the three cases involving the highest $\alpha_b$ (\textit{SmMany}, \textit{LaMany}, and \textit{BiDisp}). This shape may be approximately divided into three regimes: for $r>O(d_p)$, a slow return-to-isotropy regime occurs in which the flow relaxes towards isotropy as $r$ is decreased; a bump (seen more clearly in the semi-log plot) is observed at the scale of the bubble $r=O(d_p)$; and a third regime at $r<O(d_p)$ where the flow again relaxes towards isotropy, but at a much faster rate with decreasing $r$ than it does in the first regime. This rapid approach towards isotropy is never fully successful, however, with  significant anisotropy persisting at the smallest scales. For the \textit{SmFew} case, the first two regimes can also be identified, however, we observe scale-independent anisotropy for $r<O(d_p)$. A possible reason for this difference is that due to the low gas void fraction in the \textit{SmFew} case, scales at $r<O(d_p)$ are influenced more strongly than the other case by the single-phase behaviour, which as discussed, leads $C_{ani}$ to actually increase at the smallest scales. Moreover, we notice that the rate of return-to-isotropy is stronger with increasing $Re_p$, even though the anisotropy becomes weaker with increasing $Re_p$. This is in contrast with the phenomenological notion from single-phase homogeneous anisotropic turbulence that the rate of return is typically faster for more strongly anisotropic flows, and is related to the nature of the slow part of the pressure-strain interaction term \citep{1995_Chung}. The present observation implies, however, that in bubbly flows the additional rapid pressure-strain term arising from the bubble-induced force production \citep{2020_Ma} may play an important role in the overall return to isotropy in bubbly turbulent flows. More extensive DNS data would be required in order to investigate this in detail.

\section{High-order structure functions}\label{sec: high-order SF}

\subsection{Energy transfer and third-order structure functions}\label{subsec: 3.SF}

The third-order structure function is of particular significance since it is related to the mean nonlinear energy transfer among the scales of a turbulent flow \citep{2018_Alexakis}. In particular, in the Karman-Howarth type equation governing $D_{ii}(r_3,t)$, the inter-scale energy transfer term is \citep{2001_Hill}
\begin{equation}
    \mathcal{F}(r_3,t)=\sum_{\gamma=1}^3\mathcal{F}_\gamma(r_3,t)\equiv \sum_{\gamma=1}^3\frac{\partial}{\partial r_\gamma}D_{\gamma ii}(r_3,t),\label{eq: F}
\end{equation}
where $D_{\gamma ii}(r_3,t)\equiv \langle\Delta {u'_\gamma}(r_3,t) \Delta u'_i(r_3,t)\Delta u'_i(r_3,t)\rangle$ (here we are writing $r_3$ in the function argument as short-hand for $r_3\boldsymbol{e}_3$). When $\mathcal{F}<0$ this corresponds to the nonlinear term supplying energy to the scale, while $ \mathcal{F}>0$ corresponds to the nonlinear term removing energy from the scale. In a single-phase, three-dimensional turbulent flow, there is on average a downscale flux of energy with $\mathcal{F}<0$ in the inertial and dissipation ranges, representing the transfer of energy injected into the flow at the large-scales down to the smallest scales where it is dissipated due to viscous stresses. However, the presence of bubbles in the flow could change this picture, since bubbles produce kinetic energy at scales $O(d_p)$ and some of this may then be transferred up towards the larger scales, as well as some being transferred down towards smaller scales. It is of great interest to understand this basic question of how the bubbles modify the energy transfer mechanisms and behaviour in the turbulent flow compared with the single-phase behaviour and mechanisms of strain self-amplification and vortex stretching \citep{2020_Carbone,2020_Johnson}. However, since our DNS data is for a fixed streamwise and wall-normal location, we are only able to compute the $\mathcal{F}_3$ contribution in \eqref{eq: F}, and thereby cannot fully determine $\mathcal{F}$. Nevertheless, we can still partially explore the question of the effect of the bubbles on the energy transfer by considering $\mathcal{F}_3$.

\begin{figure}
	\begin{minipage}[b]{1.0\linewidth}
		\begin{minipage}[b]{0.5\linewidth}
			\centering
			\makebox[1.6em][l]{\raisebox{-\height}{(\textit{a})}}%
			\raisebox{-\height}{\includegraphics[height=3.7cm]{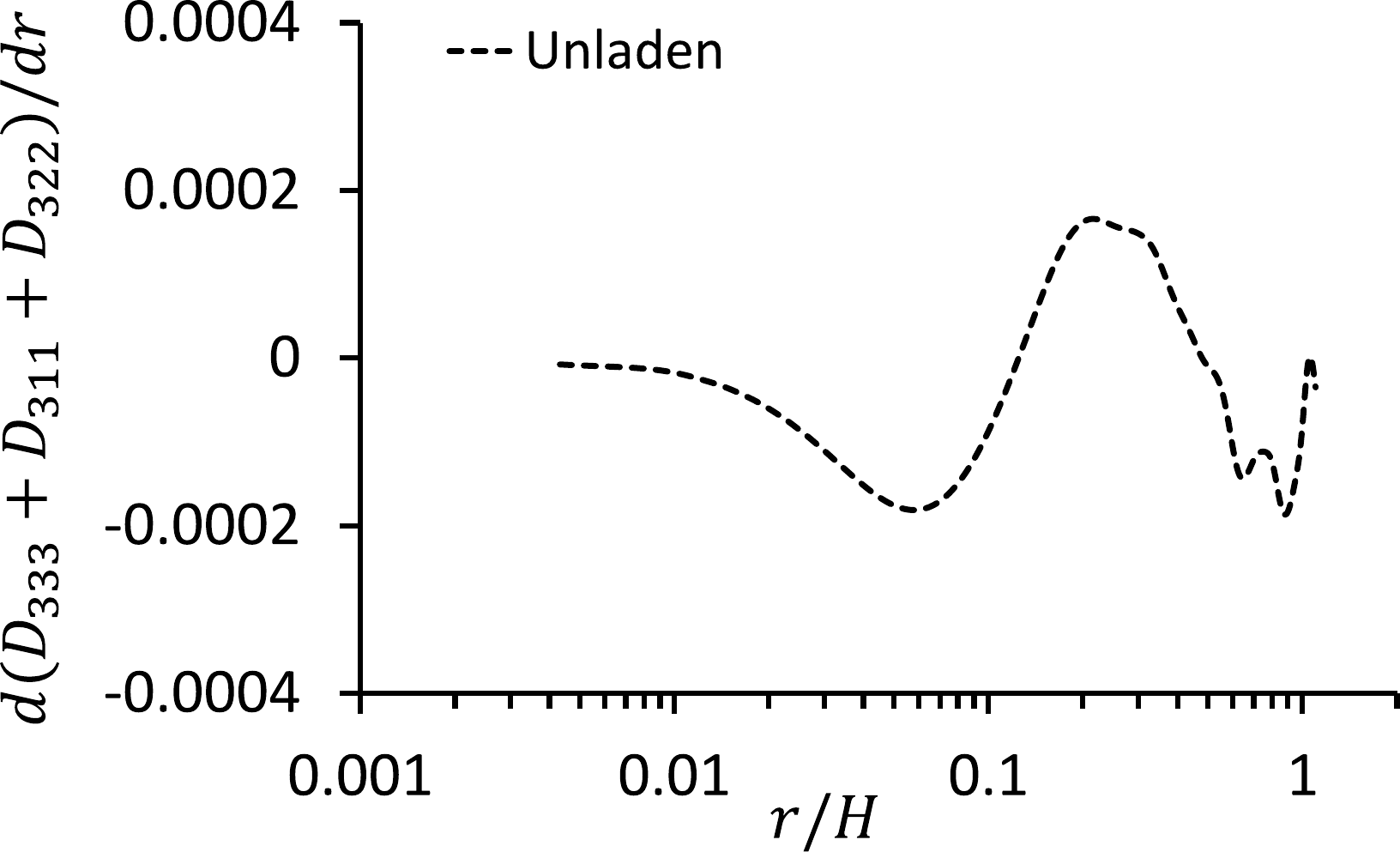}}
		\end{minipage}
		\begin{minipage}[b]{0.5\linewidth}
			\centering
			\makebox[1.6em][l]{\raisebox{-\height}{(\textit{b})}}%
			\raisebox{-\height}{\includegraphics[height=3.7cm]{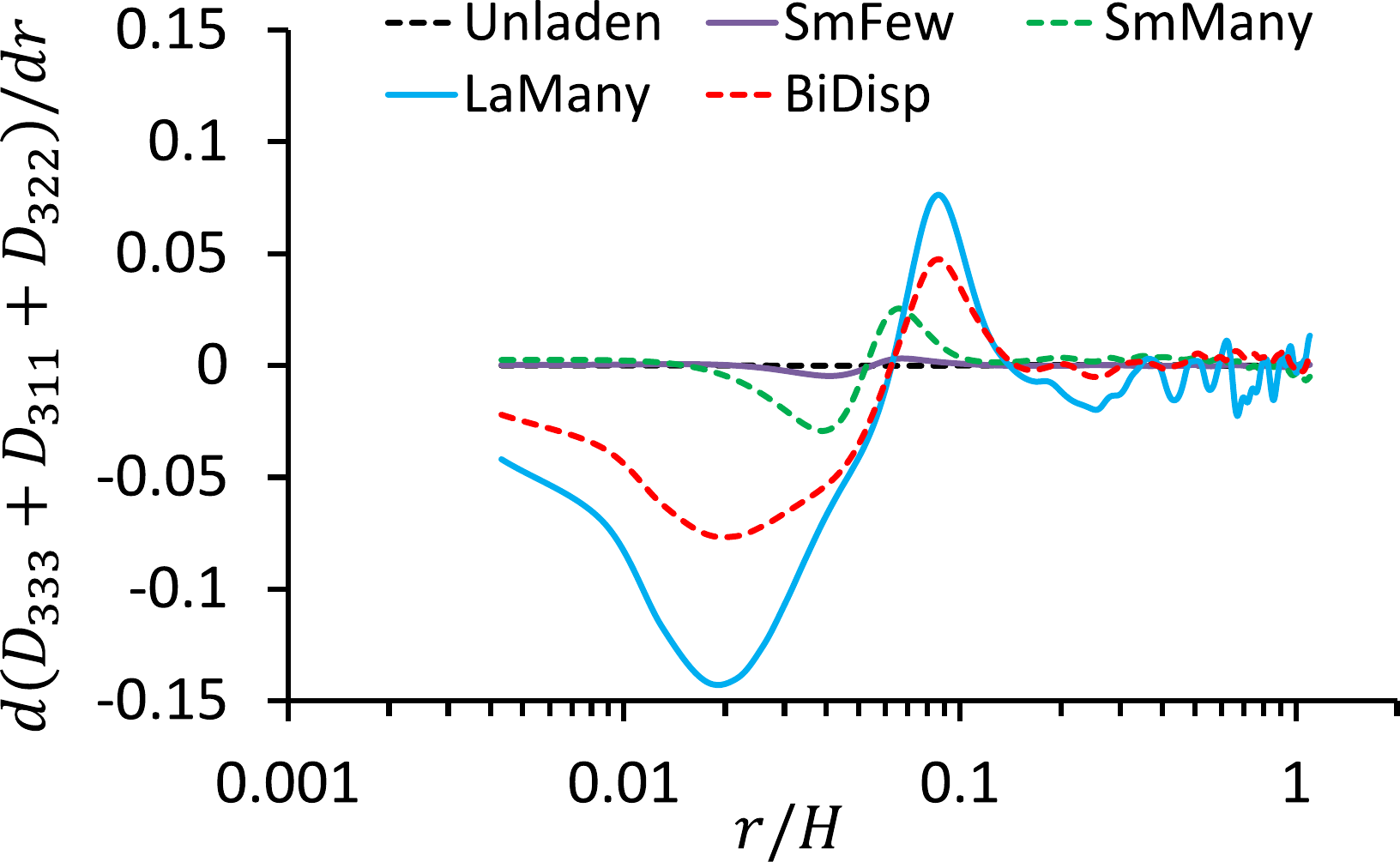}}
		\end{minipage}
	\end{minipage}
	\caption{Contribution $\mathcal{F}_3$ to the nonlinear energy transfer term: (\textit{a}) Unladen case, (\textit{b}) all the cases.} \label{fig: NET}
\end{figure}

In figure \ref{fig: NET}(\textit{a}) we plot $\mathcal{F}_3$ for the unladen case. The results show that at larger scales $\mathcal{F}_3>0$, while at smaller scales $\mathcal{F}_3<0$, corresponding to energy being taken from the large scales and passed down to the small scales via the nonlinear transfer term. At this low Reynolds number, there is no inertial range over which $\mathcal{F}_3$ is constant, indicating that there is no cascade in the strict sense. In figure \ref{fig: NET}(\textit{b}) we plot $\mathcal{F}_3$ for all the cases. The first thing that is apparent from these results is that the nonlinear energy transfer is in general much stronger for the bubble cases than the unladen case. For example, for the \textit{LaMany} case, the peak magnitude of $\mathcal{F}_3$ is three orders of magnitude larger than for the unladen case. The second thing is that the qualitative behaviour of the bubble cases is the same as the unladen case, indicating that energy is passed downscale from the large scales to the small scales in the flow. The study of \cite{2020_Pandey} explored a homogeneous bubbly turbulent flow using a Fourier space analysis and also observed that there is a downscale energy transfer, although their flow was considerably different to ours, and as noted in the introduction, there are some issues with their analysis.

Our results indicate then that the introduction of bubbles into the turbulent flow does not lead to an upscale energy transfer, at least for the cases we have considered and for the $\mathcal{F}_3$ contribution. One possible reason for this is that the energy being sent downscale from the large scales (where it is produced due to the mean-shear production), simply overwhelms any energy being transferred upscale from the bubbles, so that overall the energy transfer is downscale. However, since in most of these bubble cases the dominant source of TKE comes from bubble induced production \citep{2016_Santarelli}, rather than mean-shear production, this does not seem likely. A second possible reason is that since the scales at which the bubbles are producing kinetic energy, $O(d_p)$, are also scales at which the viscous forces in the flow are strong, then most of the kinetic energy produced by the bubbles may be simply dissipated before it is able to be transferred upscale by nonlinear forces in the flow. A third possible reason relates to the fact that although the bubble length scale is $d_p$, they may inject energy at scales significantly larger than this. Indeed, the wakes produced by bubbles can have a length that is $O(10 d_p)$. For the Reynolds number of our DNS the scale separation between the bubble diameter and the large scales of the flow is not very large, and therefore the bubbles may actually directly inject TKE into scales on the order of the large scales of the flow, and this energy is then able to be sent downscale via the nonlinear energy transfer term. If this third explanation is true, then an interesting implication is that if BIT flows can develop with $d_p\lll \ell_I$ (where $\ell_I$ is the integral length scale of the flow), then the maximum scale at which the bubbles will inject energy into the flow will be much smaller than $\ell_I$. In this case, an upscale energy transfer must occur at scales $\gg d_p$ since otherwise the large scales would have no source of energy.

\begin{figure}
	\centering
	\includegraphics[height=3.8cm]{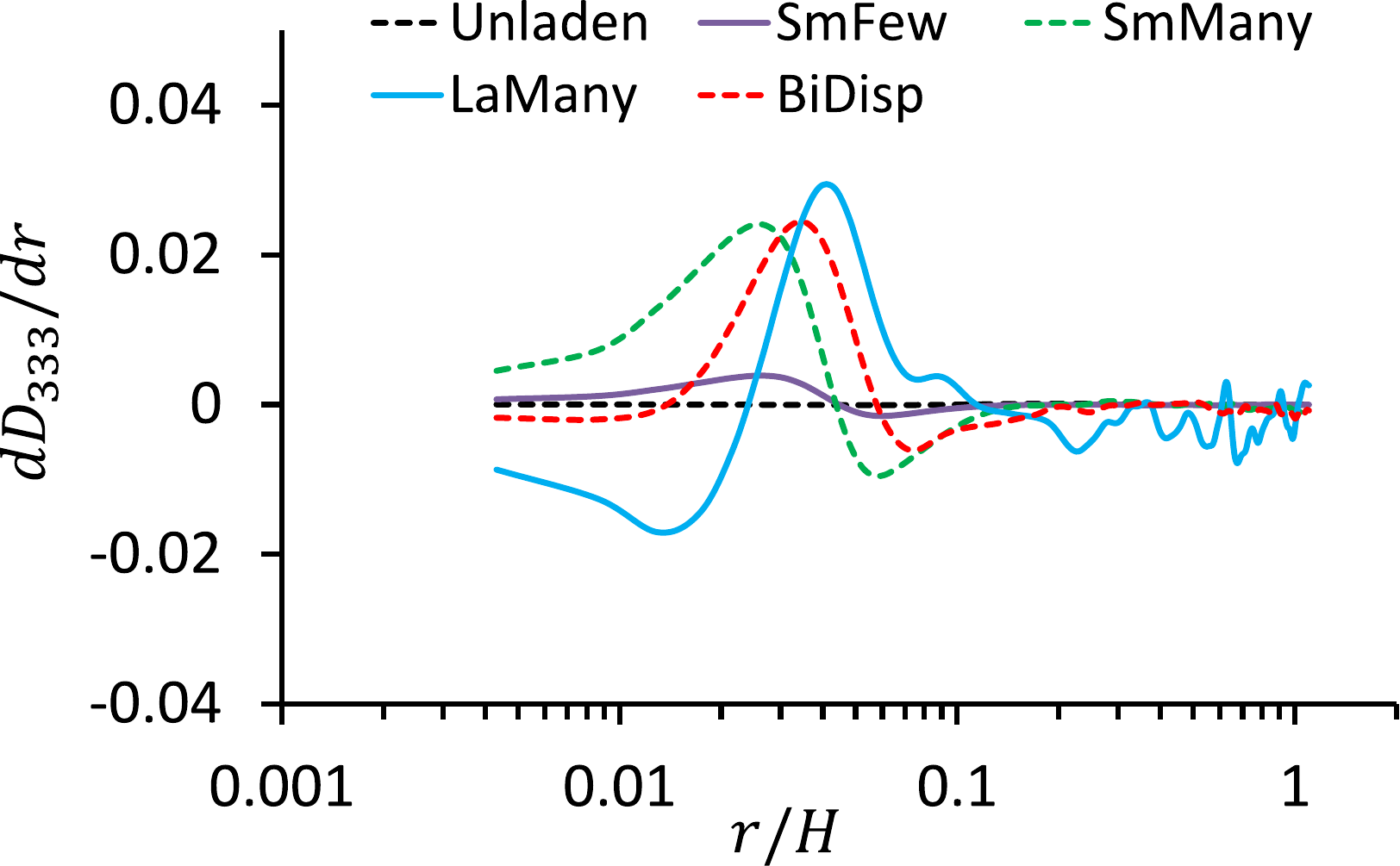}
	\caption{Derivative of $D_{333}$, the third-order longitudinal structure function.} \label{fig: derivative_D333}
\end{figure}

In figure \ref{fig: derivative_D333} we plot one of the contributions to $\mathcal{F}_3$, namely $d D_{333}/dr$. In contrast to figure \ref{fig: NET}, this result shows that for the \textit{SmFew} and \textit{SmMany} cases there is an upscale transfer of energy, with energy being extracted from the small scales and passed to the large scales. This reveals a subtle effect of the bubbles; while overall the bubble cases still exhibit a downscale energy transfer, the energy transfer associated with particular components of the velocity field do exhibit an upscale energy transfer.

\begin{figure}
	\begin{minipage}[b]{1.0\linewidth}
		\begin{minipage}[b]{0.5\linewidth}
			\centering
			\makebox[0.5em][l]{\raisebox{-\height}{(\textit{a})}}%
			\raisebox{-\height}{\includegraphics[height=3.6cm]{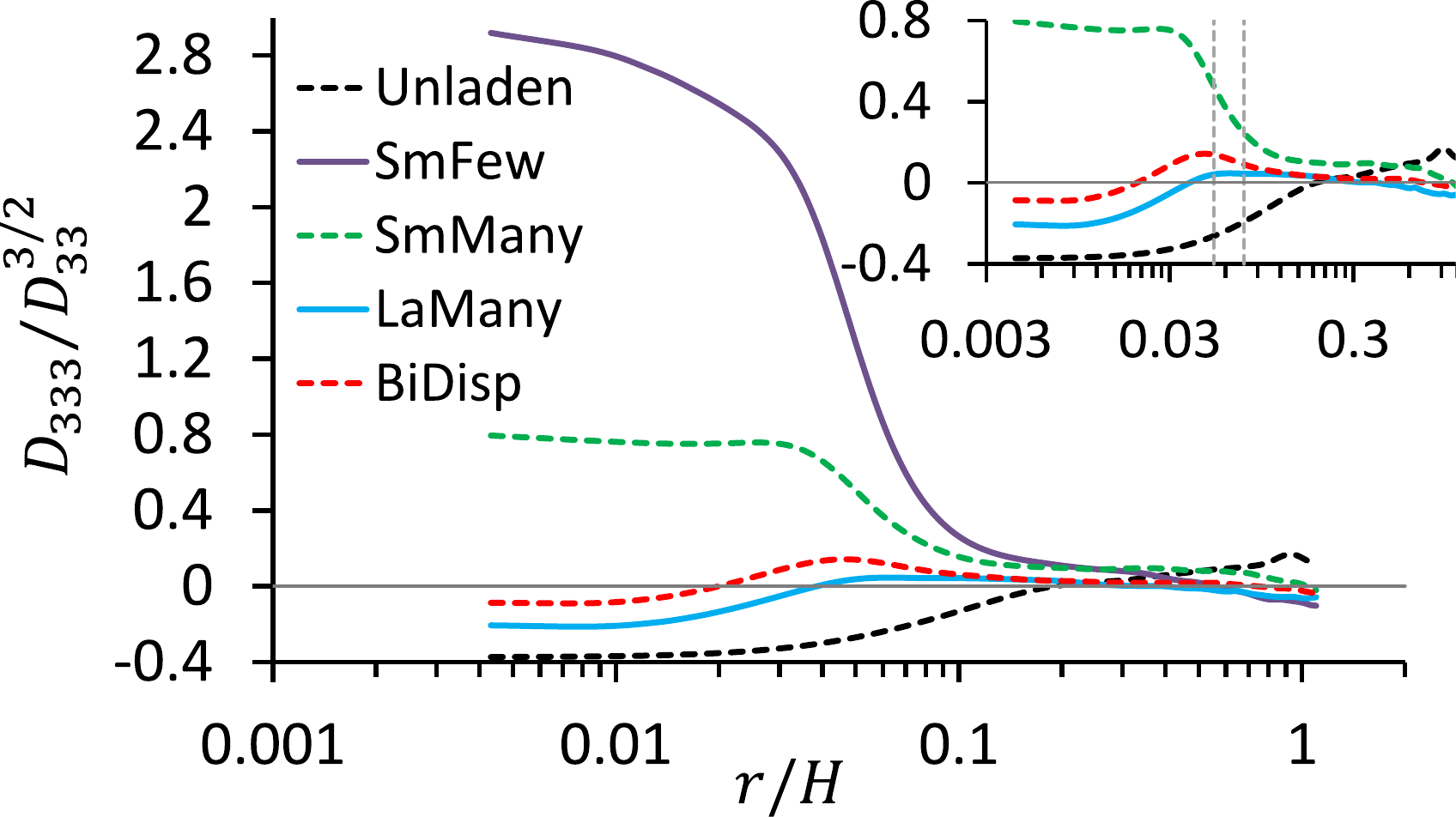}}
		\end{minipage}
		\begin{minipage}[b]{0.5\linewidth}
			\centering
			\makebox[0.5em][l]{\raisebox{-\height}{(\textit{b})}}%
			\raisebox{-\height}{\includegraphics[height=3.6cm]{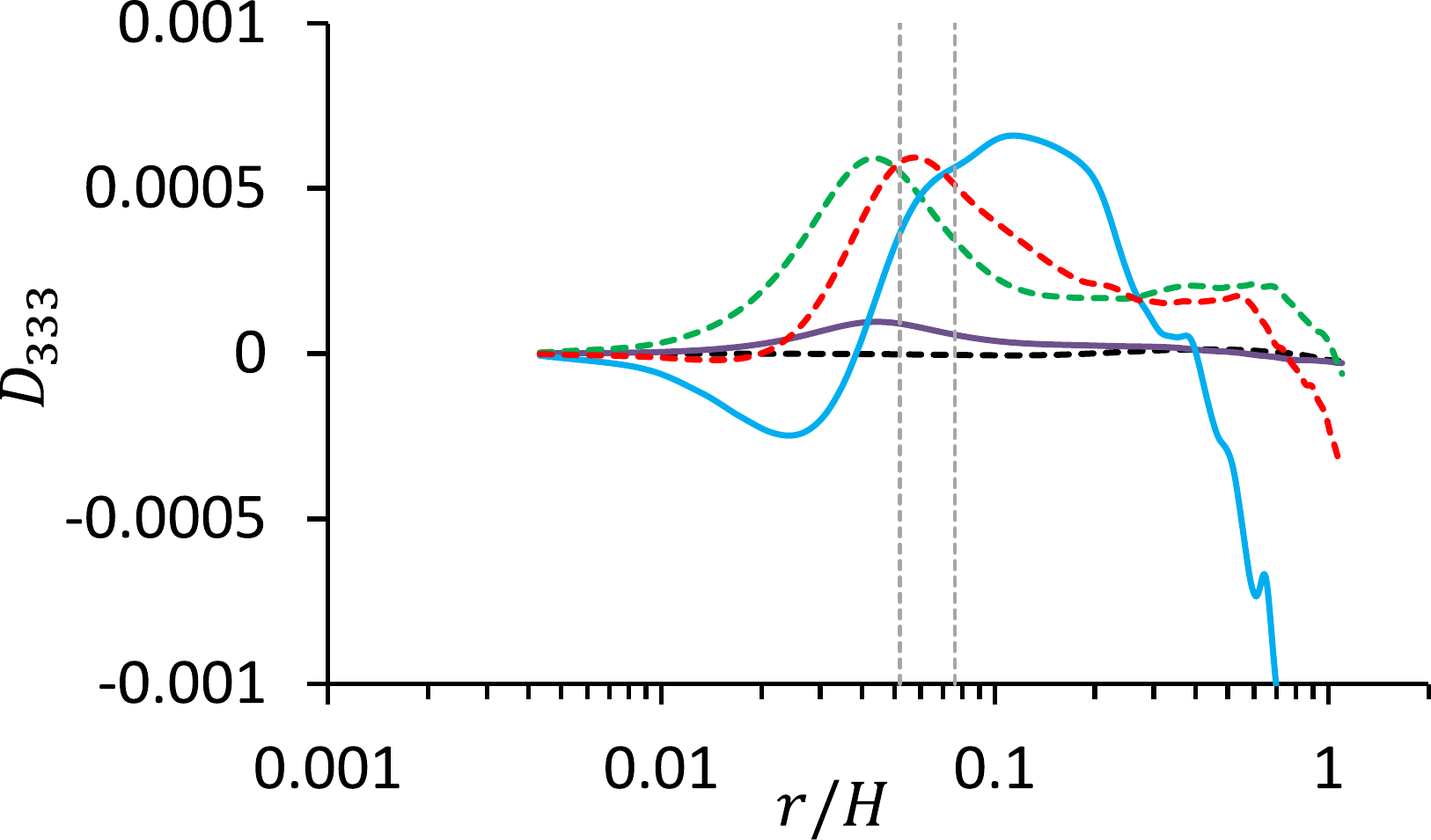}}
		\end{minipage}
	\end{minipage}
	\caption{Third-order longitudinal structure functions: (\textit{a}) Skewness; (\textit{b}) Unnormalized. The inset in (\textit{a}) shows zoomed for the cases excluding \textit{SmFew}. The two vertical dashed lines in both the inset of (\textit{a}) and in (\textit{b}) show $r=d_p$ for smaller and larger bubbles, respectively. The horizontal line in (\textit{a}) is the value of $0$.} \label{fig: D333}
\end{figure}

Figure \ref{fig: D333} shows $D_{333}/D_{33}^{3/2}$ (the skewness of the longitudinal velocity increment), as well as the unnormalized form $D_{333}$. For the present unladen case, in the channel centre we find $D_{333}/D_{33}^{3/2}\approx-0.4$ for $r\rightarrow0$, which is in close agreement with the value found in isotropic turbulence \citep{Ishihara_2007,2015_Davidson}. The Reynolds number of the channel flow is too low for there to be a clear inertial range, and therefore the unladen data for $D_{333}$ does not show evidence of the \textquotedblleft Fourth-Fifths law\textquotedblright\ $D_{333}=-(4/5)\langle\epsilon\rangle r$ predicted for the inertial range of isotropic turbulence by \cite{1941_Kolmogorov_b}. For the cases involving bubbles, a first observation is that $D_{333}/D_{33}^{3/2}$ is positive for all cases at $r=O(d_p)$, and for the \textit{SmFew} and \textit{SmMany} cases it remains positive in the limit $r\to 0$, while for others it becomes negative in this limit. The data for $D_{333}$ indicates that the fluctuations are much larger for the bubble cases than for the unladen case, especially those with higher volume fractions, while the results for the skewness shows that the bubble cases have a skewness that is either larger or smaller in magnitude than the unladen case, depending upon the case. The \textit{SmFew} case is particularly interesting, since it exhibits a very large value of the skewness for $r<O(d_p)$, even though this case has the smallest volume fraction $\alpha_b$. However, as expected, the results for $D_{333}$ for this case reveal that the magnitude of $D_{333}$ is much smaller for the \textit{SmFew} case than for the other three bubble cases that have higher values of $\alpha_b$. This shows how sensitive certain properties of the turbulent flow are to the presence of the bubbles, even for relatively low $\alpha_b$. It also implies that the bubbles could have a significant effect on tracer particle dispersion in turbulent flows, whose irreversibility is intimately related to the asymmetry in the velocity increment distributions \citep{2014_Jucha,2016_Bragg,2017_Bragg}.

The transverse structure functions $D_{111}$ and $D_{222}$ were also considered for each case. The associated skewness values were found to be very close to zero for all cases (not shown here). 

\subsection{Fourth and sixth-order structure functions}\label{subsec: 4.6.8.SF}

\begin{figure}
	\begin{minipage}[b]{1.0\linewidth}
		\begin{minipage}[b]{0.5\linewidth}
			\centering
			\makebox[0.5em][l]{\raisebox{-\height}{(\textit{a})}}%
			\raisebox{-\height}{\includegraphics[height=3.6cm]{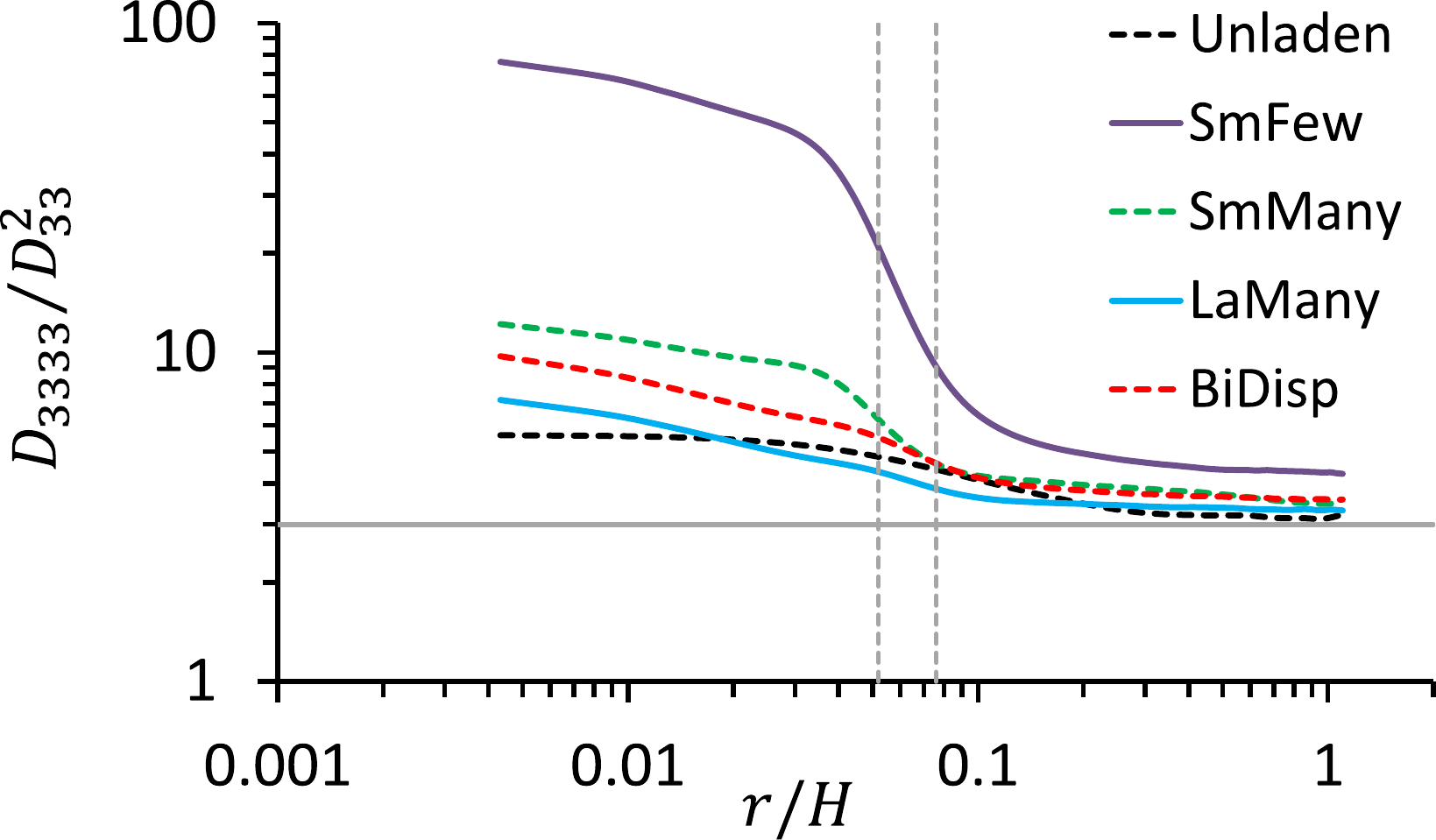}}
		\end{minipage}
		\begin{minipage}[b]{0.5\linewidth}
			\centering
			\makebox[0.5em][l]{\raisebox{-\height}{(\textit{b})}}%
			\raisebox{-\height}{\includegraphics[height=3.6cm]{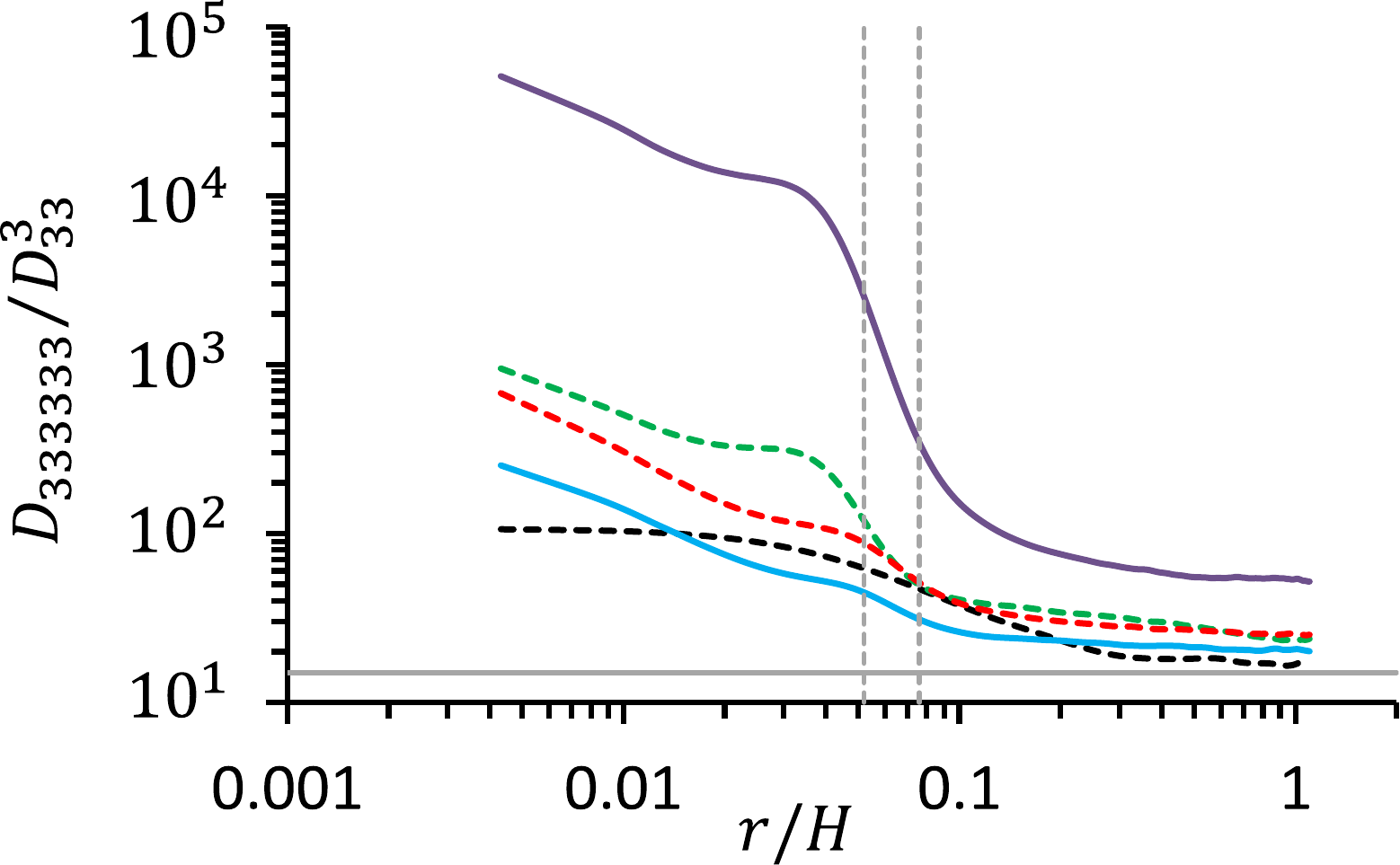}}
		\end{minipage}
	\end{minipage}
	\begin{minipage}[b]{1.0\linewidth}
		\vspace{3mm}
		\begin{minipage}[b]{0.5\linewidth}
			\centering
			\makebox[0.5em][l]{\raisebox{-\height}{(\textit{c})}}%
			\raisebox{-\height}{\includegraphics[height=3.6cm]{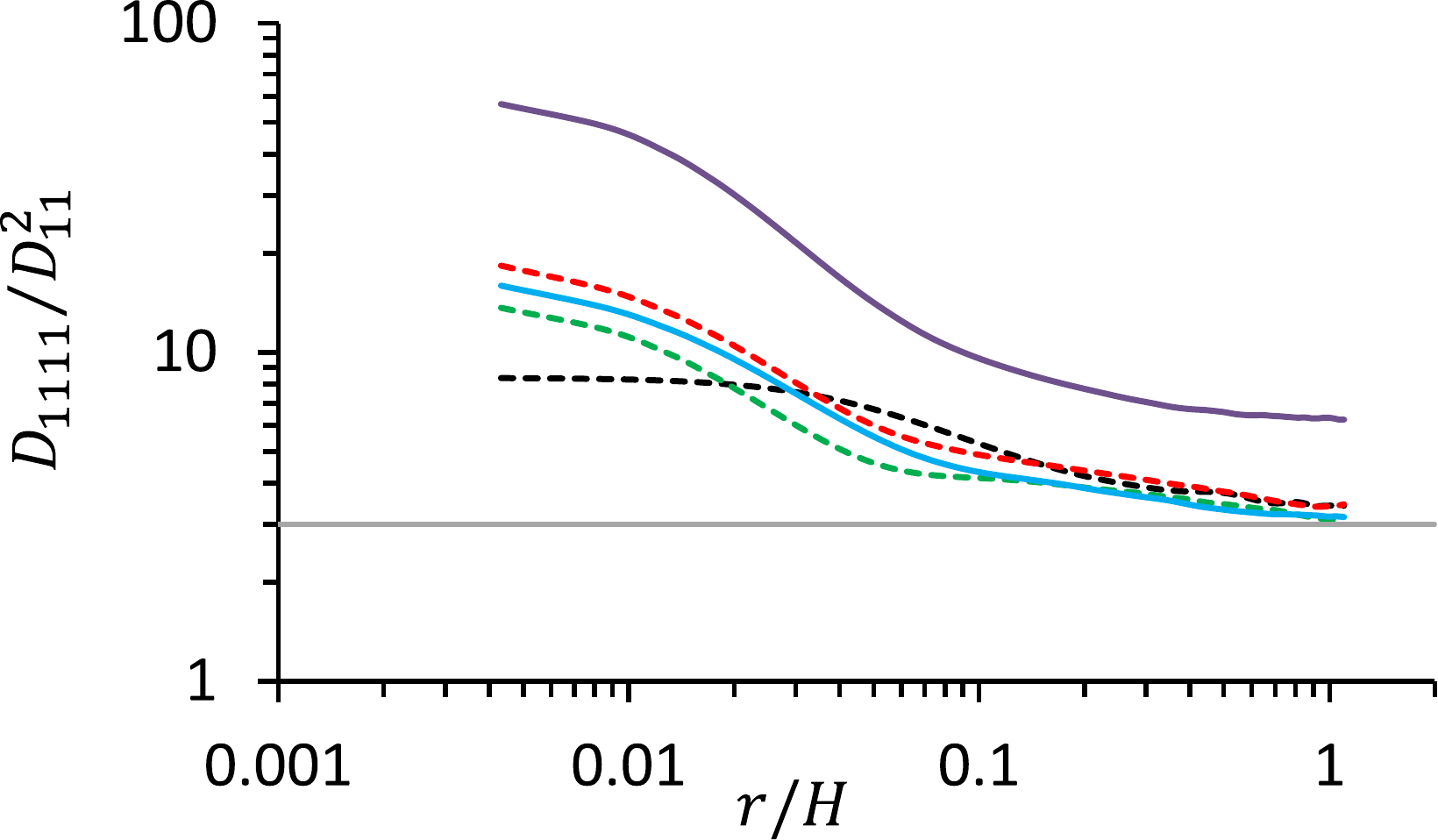}}
		\end{minipage}
		\begin{minipage}[b]{0.5\linewidth}
			\centering
			\makebox[0.5em][l]{\raisebox{-\height}{(\textit{d})}}%
			\raisebox{-\height}{\includegraphics[height=3.6cm]{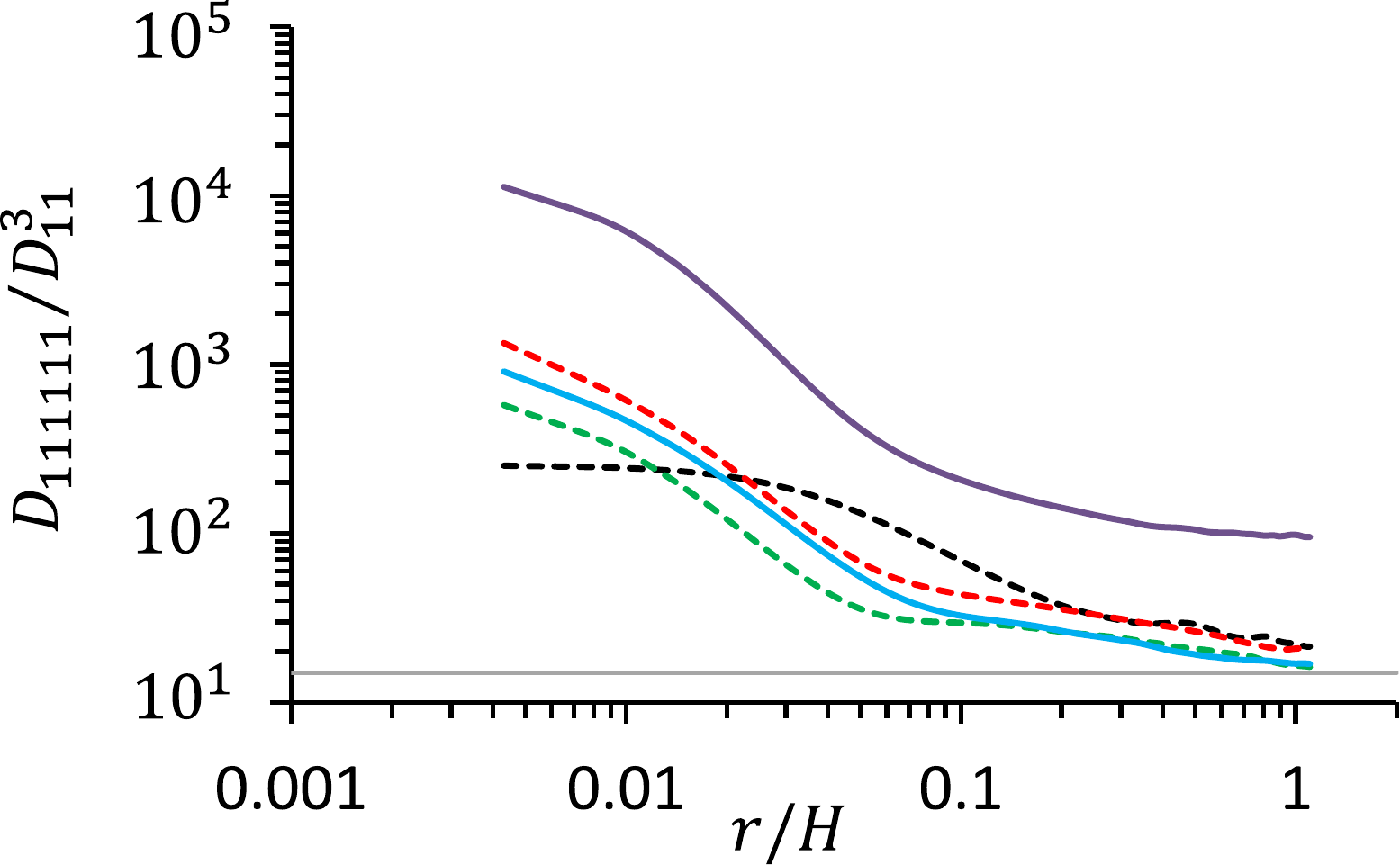}}
		\end{minipage}
	\end{minipage}
	\begin{minipage}[b]{1.0\linewidth}
		\vspace{3mm}
		\begin{minipage}[b]{0.5\linewidth}
			\centering
			\makebox[0.5em][l]{\raisebox{-\height}{(\textit{e})}}%
			\raisebox{-\height}{\includegraphics[height=3.6cm]{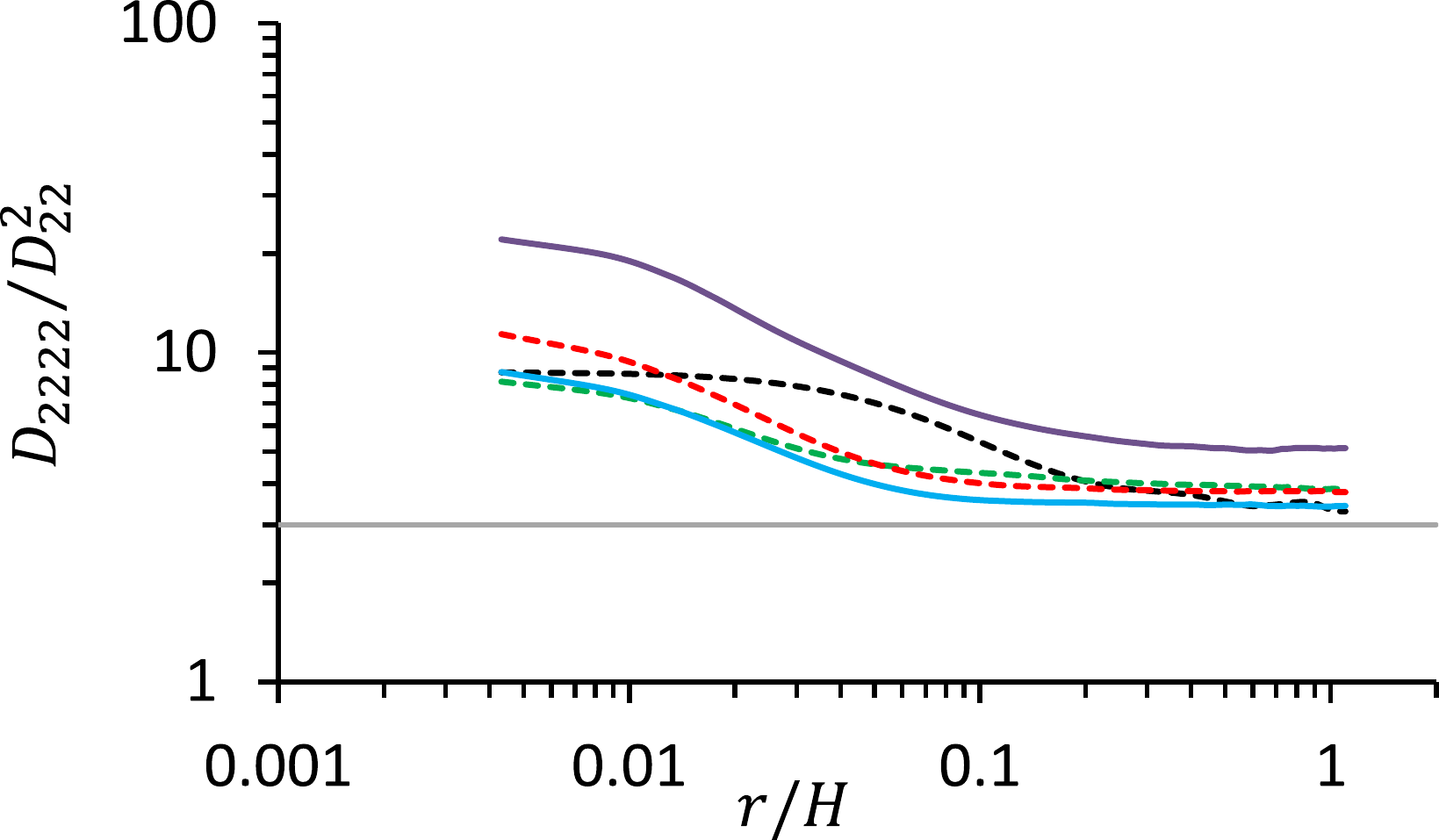}}
		\end{minipage}
		\begin{minipage}[b]{0.5\linewidth}
			\centering
			\makebox[0.5em][l]{\raisebox{-\height}{(\textit{f})}}%
			\raisebox{-\height}{\includegraphics[height=3.6cm]{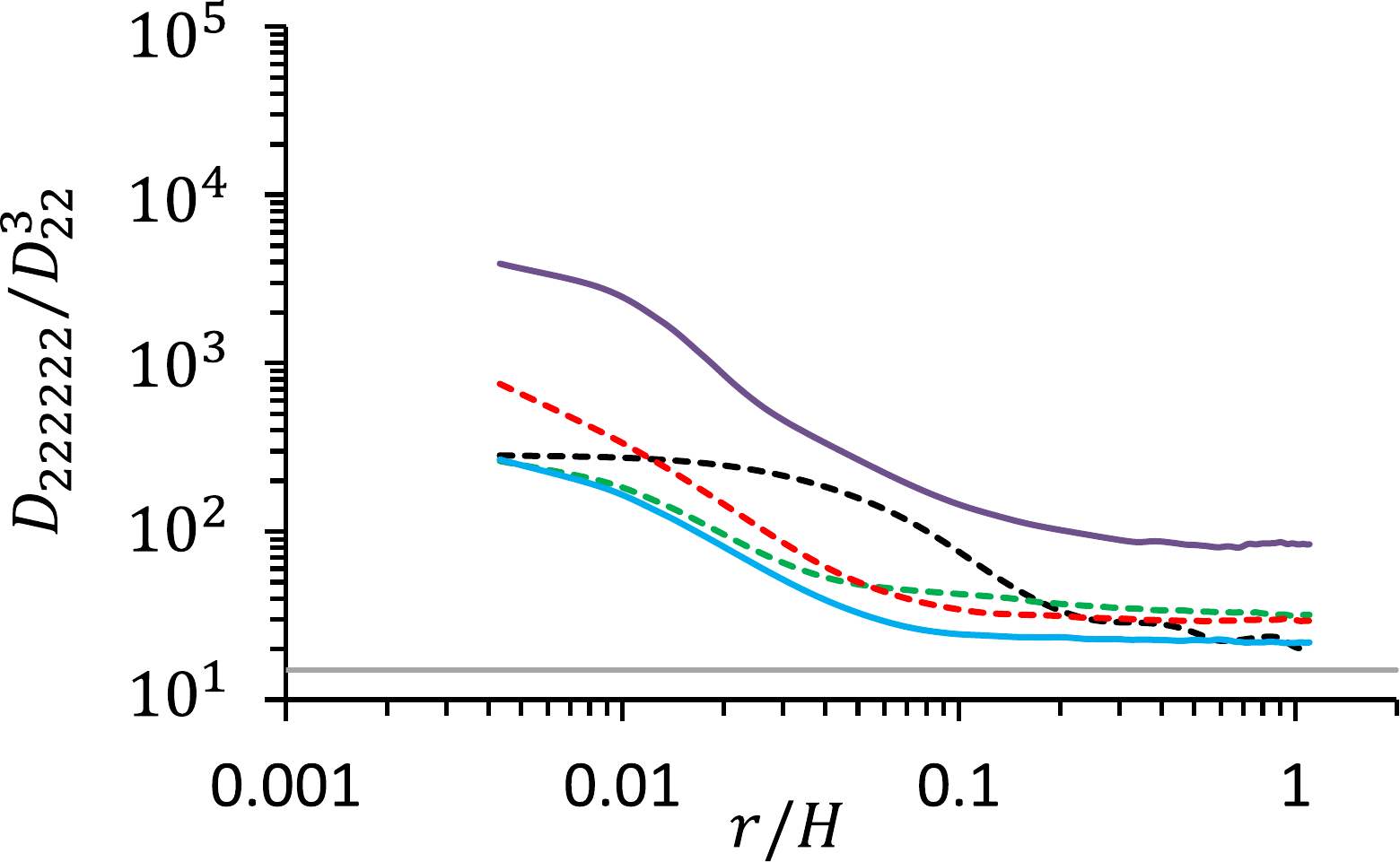}}
		\end{minipage}
	\end{minipage} 
	\caption{Normalized fourth- (\textit{a,c,e}) and sixth-order (\textit{b,d,f}) structure functions, with (\textit{a,b}) the longitudinal ones and (\textit{c,d,e,f}) the transverse ones. The two vertical dashed lines in (\textit{a,b}) show $r=d_p$ for smaller and larger bubbles, respectively. The horizontal lines in (\textit{a,c,e}) and in (\textit{b,d,f}) are the Gaussian values of $3$ and $15$ for flatness and superflatness, respectively.}\label{fig: flatness}
\end{figure}

We now consider the fourth and sixth-order structure functions to explore how the bubbles influence the intermittency of the background turbulence. In figures \ref{fig: flatness}(\textit{a,c,e}) we consider the normalized fourth-order moments (i.e. the flatness of the velocity increments) in each of the directions, denoted by $D_{1111}/D_{11}^{2}$, $D_{2222}/D_{22}^{2}$, and $D_{3333}/D_{33}^{2}$. If the velocity increments had Gaussian probability distributions, then these quantities would be equal to three at all scales. Departures from three indicate non-Gaussianity in the velocity increment distributions, while spatial dependence of these quantities indicates that the velocity increments are not only non-Gaussian, but also intermittent \citep{1995_Frisch}.

For the unladen case, the results in figures \ref{fig: flatness}(\textit{a,c,e}) show that at the large scales, the quantities are close to three, but increase as $r$ decreases. Furthermore, the results show that the departures from Gaussianity are stronger for the transverse components than they are for the longitudinal component, also observed in isotropic turbulence \citep{2009_Ishihara}. The effect of the bubbles on these quantities is non-trivial. At the very smallest scales, the bubble cases show much stronger non-Gaussianity and intermittency in the turbulent flow than for the unladen case. A similar finding was also observed in the experiment of \cite{2005_Rensen} for their fully developed turbulent bubbly flow and in \cite{2012_Biferale} when comparing the small-scale properties of boiling and non-boiling convective turbulent flows. However, with the exception of the \textit{SmFew} case, the bubbles tend to suppress intermittency in the flow at scales comparable to their diameter, for the streamwise and wall-normal velocity directions. Roughly speaking, it is only at scales smaller than $O(d_p)$ that the flatness starts to increase significantly.

The flatness results for the \textit{SmFew} case reveal much stronger non-Gaussianity and intermittency in the flow compared to the other three bubble cases that have much higher gas void fraction ($\approx 7$ times than \textit{SmFew}). Some insight into this behaviour can be found by considering the ways in which the bubbles modify the turbulent flow. The most significant qualitative modification the bubbles make to the turbulent flow is the wakes they generate, and the flow properties in these wakes will typically be significantly different from the properties of the background turbulence. When the number of bubbles is not too large (but still large enough to make a statistically significant contribution to the flow properties), there are relatively few regions in the flow where the flow differs due to the wakes from everywhere else (i.e. these regions are \textquotedblleft rare\textquotedblright), breaking self-similarity in the flow and enhancing intermittency compared to the case without the bubbles. As the number of bubbles increases, the regions of the flow occupied by the wakes becomes less rare, and hence their contribution to the intermittency reduces. Moreover, as the number of bubbles/size of the bubbles increases, the possibility for wake interaction increases, and this can enhance mixing in the flow, which would reduce the impact of the wakes on the flow intermittency.

The behaviour observed for the flatness is also observed for the sixth- (superflatness, see figures \ref{fig: flatness}\textit{b,d,f}) and eighth-order (hyperflatness, not shown here) structure functions. A most surprising result concerns the relationship between $Re_H$ and the level of intermittency. For example, as shown in figure \ref{fig: Re_H}, the \textit{LaMany} case has a much larger $Re_H$ than the \textit{SmFew} case, yet the latter exhibits much stronger intermittency in the flow than the former. This is in striking contrast to single-phase turbulence where small-scale intermittency increases with increasing Reynolds number \citep{1995_Frisch}. This indicates that the mechanisms generating intermittency in bubble-laden turbulent flows might be significantly different from those in single-phase turbulence. Further data would be required to explore the mechanisms responsible for this behaviour.

\section{Conclusions}

In this paper we have analyzed various properties of bubble-laden turbulent flows across the range of scales of the flow, including the anisotropy, energy transfer, and intermittency in the flow.

To explore the anisotropy of the flow at different scales, we developed an extension of the barycentric map approach that was previously developed for analyzing the one-point Reynolds stress tensor \citep{2007_Banerjee}. In our approach, the anisotropy at any scale may be quantified and visualized, as well as providing information on the componentiality of the flow at the scale. Using this we were able to explore how the bubbles modify anisotropy in the flow. We found that the bubbles significantly enhance anisotropy in the flow at all scales compared with the unladen case, and that for some bubble cases, very strong anisotropy persists down to the smallest scales of the flow. The strongest anisotropy was observed for the cases involving small bubbles. Deviations from the behaviour that would be expected for an isotropic flow occur not only in the streamwise/gravity direction, but also in the other directions.

Concerning the energy transfer among the scales of the flow, our DNS data did not allow us to thoroughly explore this, but we were able to consider a number of its important aspects. Our results indicate that for the bubble-laden cases, the energy transfer is from large to small scales, just as for the unladen case. However, there is evidence of an upscale transfer when considering the transfer of energy of particular components of the velocity field, rather than the full kinetic energy involving all three components of the flow. We also conjectured, however, that in a bubble-laden channel flow at sufficiently high Reynolds number, and with channel height sufficiently large compared with the bubble size, there may exists an upscale transfer of energy at scales much larger than the bubble diameter. Our DNS does not lie in the parameter range required to observe such behaviour. Another important finding is that although the direction of the energy transfer is the same with and without the bubbles, the energy transfer is much stronger (by several orders of magnitude) for the bubble-laden cases, suggesting that the bubbles play a strong role in enhancing the activity of the nonlinear term in turbulent flows.

We also considered the normalized forms of the fourth and sixth-order structure functions, corresponding to the flatness and superflatness of the fluid velocity increments. These results revealed that the introduction of bubbles into the flow strongly enhances the intermittency of the turbulence in the dissipation range, but suppresses it at scales comparable to the bubble diameter. The \textit{SmFew} case, however, shows enhancements of intermittency at all scales. We interpreted the effect of the bubbles on the flow intermittency in terms of the contributions of the bubble wakes to the overall properties of the turbulence. This strong enhancement of the dissipation scale intermittency has significant implications for understanding how the bubbles might modify the mixing properties of turbulent flows, and the associated large velocity gradients. These will be the subject of future investigations.

\section*{Acknowledgements}

The authors would like to acknowledge Michele Iovieno for providing some of his DNS data to validate the code used in the present study. T.M. benefited from discussions on this topic with Peter Brugger, Maurizio Carbone and acknowledges funding by Deutsche Forschungsgemeinschaft (DFG, German Research Foundation) under Grants MA 8408/1-1 and MA 8408/2-1.\\

\section*{Declaration of Interests}
The authors report no conflict of interest.

\bibliographystyle{jfm}
\bibliography{Aniso_BIT}

\end{document}